\pdfoutput=1

\documentclass[a4paper,11pt]{article}
\usepackage[colorlinks=true, urlcolor=citecolor, linkcolor=citecolor, citecolor=citecolor]{hyperref}

\usepackage{bm}
\usepackage{latexsym}
\usepackage{amsmath}
\usepackage{amsfonts}
\usepackage{amssymb}
\usepackage{amsthm}
\usepackage{booktabs}
\usepackage{enumerate}
\usepackage{enumitem}
\usepackage{float}
\usepackage[hmargin = 1.6cm, vmargin = 3cm]{geometry}
\usepackage{graphicx}
\usepackage[normalem]{ulem}
\usepackage{multirow}
\usepackage{natbib}

\def\T{{ \mathrm{\scriptscriptstyle T} }}
\DeclareMathOperator{\iid}{\overset{iid}{\sim}}
\def\bmx{x}
\def\Exp{\text{Exp}}
\date{}
\begin{document}

\title{\vspace{-1cm}Affinity-based measures of medical diagnostic test accuracy}

\author{Miguel~de Carvalho\\[4pt]
\textit{School of Mathematics, University of Edinburgh, EH 91PQ, UK}
\\[2pt]
{\texttt{miguel.decarvalho@ed.ac.uk}}\\ [4pt]
Bradley Barney \quad Garritt Page
\\[4pt]
\textit{Department of Statistics, Brigham Young University, UT 84602, USA}}

\markboth%
{M.~de Carvalho and others}
{Affinity-based measures of diagnostic test accuracy}

\maketitle

\def\dif{\mathrm{d}}


\def\tint{\mathop{\textstyle \int}}%
\def\ROC{\text{ROC}}
\def\AUC{\text{AUC}}
\def\YI{\text{YI}}
\def\DP{\text{DP}}
\def\betab{\beta}
\def\thetab{\theta}

\newcommand{\D}{\ensuremath{\bar{D}}}

\newcommand{\sqrtfd}{\ensuremath{\sqrt{f_D}}}
\newcommand{\sqrtfdw}{\ensuremath{\sqrt{f_D^{\omega}}}}
\newcommand{\sqrtfnd}{\ensuremath{\sqrt{f_{\D}}}}
\newcommand{\sqrtfndw}{\ensuremath{\sqrt{f_{\D}^{\omega}}}}

\newcommand{\Sqrtfd}{\ensuremath{\sqrt{f_D(y)}}}
\newcommand{\Sqrtfdw}{\ensuremath{\sqrt{f_D^{\omega}(y)}}}
\newcommand{\Sqrtfnd}{\ensuremath{\sqrt{f_{\D}(y)}}}
\newcommand{\Sqrtfndw}{\ensuremath{\sqrt{f_{\D}^{\omega}(y)}}}
\newtheorem{proposition}{Proposition}
\newtheorem{theorem}{Theorem}
\newtheorem{example}{Example}



\begin{abstract}
{\small We propose new summary measures of diagnostic test accuracy which can be used as companions to existing diagnostic accuracy measures. Conceptually, our summary measures are tantamount to the so-called Hellinger affinity and we show that they can be regarded as measures of agreement constructed from similar geometrical principles as Pearson correlation. A covariate-specific version of our summary index is developed, which can be used to assess the discrimination performance of a diagnostic test,  conditionally on the value of a predictor. Nonparametric Bayes estimators for the proposed indexes are devised, theoretical properties of the corresponding priors are derived, and the performance of our methods is assessed through a simulation study. Data from a prostate cancer diagnosis study are used to illustrate our methods.} \\

\end{abstract}
\vspace{-.3cm}
\section{Introduction}\label{sec1}
\begin{figure}
  \centering
  \includegraphics[scale = 0.45]{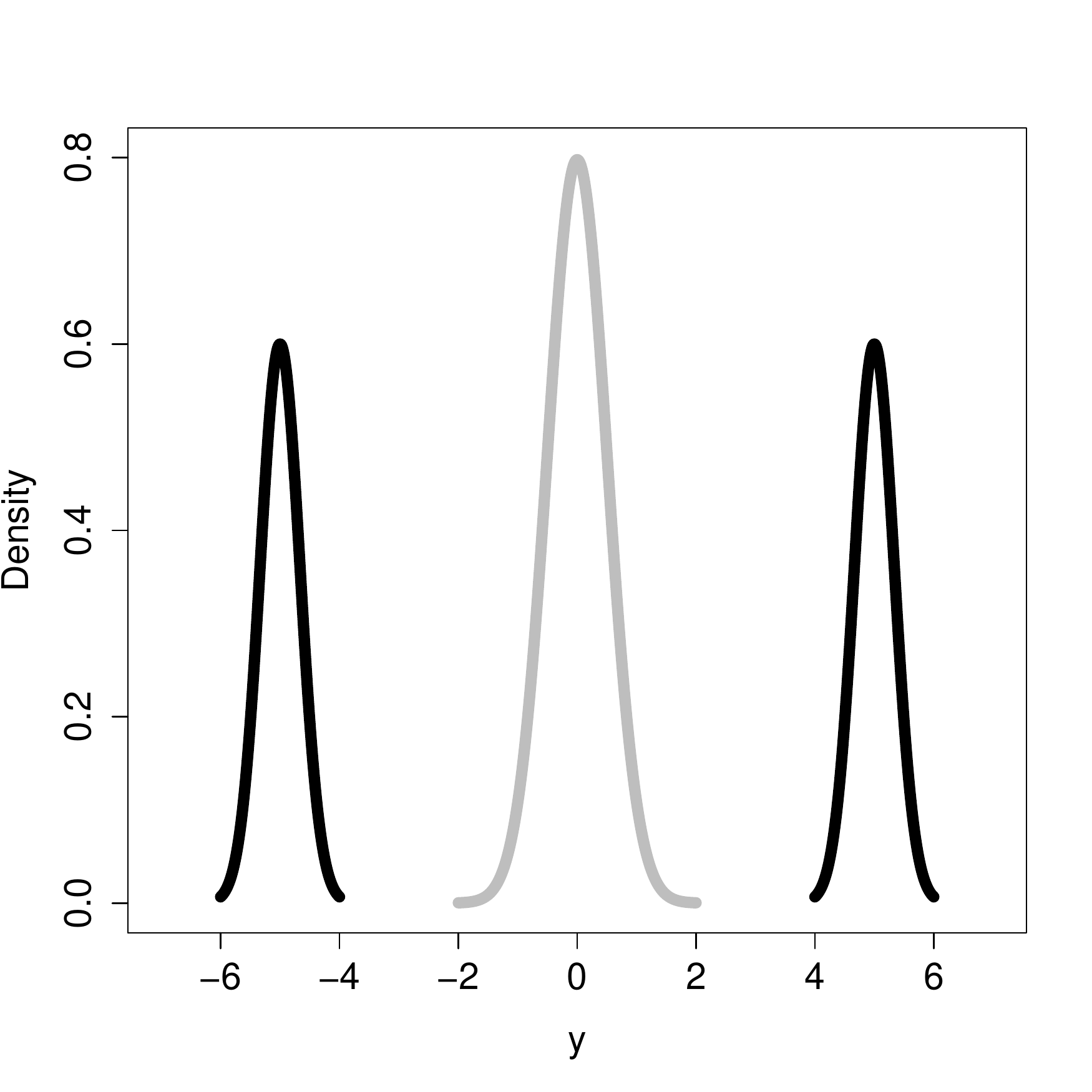}
  \caption{\small The separation trap: Perfect discrimination but $\AUC = 0.5$. Details on the truncated normal densities used to construct this example can be found in Example~\ref{trap}; the black and grey  lines respectively denote the densities of the biomarkers of the diseased and non-diseased subjects.}
  \label{motivation}
\end{figure}

Accurate diagnosis is a key target of diagnostic decision-making. Before a medical diagnostic test is routinely applied in practice, it is important to evaluate its performance in discriminating between diseased and non-diseased subjects. The most well-known summary accuracy measures are the $\AUC$ (Area Under the receiver operating characteristic Curve) and the Youden index \citep{youden1950}; other summary indexes can be found in \citet[][Section~4.3.3]{pepe2003}. {These well-known summary measures at times gloss over important differences between diseased and non-diseased subjects.  To see when this might happen let} $Y_D \sim F_D$ and $Y_{\D} \sim F_{\bar{D}}$ denote the test results for diseased and non-diseased subjects. Formally, the AUC consists of $P(Y_D > Y_{\D})$ and it is typically argued that $\AUC = 0.5$ for a test {that does} no better than chance {in discriminating between diseased and non-diseased individuals}, while $\AUC = 1$ for a test that perfectly distinguishes between diseased and non-diseased subjects. While AUC is widely used in practice, Figure~\ref{motivation} illustrates a setting under which the AUC is known to perform poorly.   {Regarding this setting, \cite{lee1996}  make the following comment}:

\begin{quote}
   ``For the two populations of the diseased and the non-diseased [...] the marker perfectly separates the two. Therefore, any clinician (or epidemiologist) will have no trouble in choosing a decision rule for the marker, that is, high and low cutoff points. Nevertheless, adopting the AUC as the measure of overall performance leads one to conclude that the marker is not better than flipping a
 fair coin (its AUC is 0.5) ''
\end{quote}

Throughout, we will refer to the situation in Figure~\ref{motivation} as the `separation trap,' since one has perfect discrimination but $\AUC = 0.5$. 
As can be seen from Figure~\ref{motivation}, even though the populations of diseased and non-diseased subjects are perfectly separated, half of the diseased-subjects are predicted to have a test result higher than the non-diseased subjects, and thus $\AUC = 0.5$. 

{But beyond the AUC, the Youden index (YI) also falls into the separation trap. To see this recall that $\protect{\text{YI}= \max_{c \in \mathbb{R}}\,\{F_{\bar{D}}(c)-F_{D}(c)\}}$, with $\text{YI}=0$ corresponding to complete overlap ($F_{\bar{D}}(c)=F_{D}(c)$), and it is often argued that $\text{YI}=1$ when the distributions are `completely separated'. It is straightforward to show that in the example in Figure~1, it holds that $\text{YI} = 1/2$, while the distributions of the markers for diseased and non-diseased subjects are completely separated---thus confirming that the Youden index would fall into the separation trap. The optimal cutoff region yielded through the Youden index} is
\begin{eqnarray*}\label{yi}
\arg \max_{c \in \mathbb{R}}\,\{F_{\bar{D}}(c)-F_{D}(c)\} = [2, 4].
\end{eqnarray*}
Interestingly, however, the more sensible cutoff region $[-4, -2] \cup [2, 4]$ could have been obtained by adjusting the definition of Youden index to consider the absolute value of the difference between distribution functions, but even this modified index would be equal to 1/2.

A main goal of this article is {to propose} new diagnostic accuracy measures {that accommodate the separation trap and that} can be used as companions, or possibly as alternatives, to existing diagnostic accuracy measures. {Conceptually, our summary measures can be motivated by first considering a geometric interpretation of covariance and Pearson correlation}.  By recalling the well-known fact that for zero-mean finite-variance random variables $X$ and $Y$, the covariance can be interpreted as an inner product between random variables \citep{williams1991}, it follows that Pearson correlation 
\begin{equation}
  \label{eq:pearson}
  \rho = \frac{\text{cov}(X, Y)}{\text{sd}(X) \text{sd}(Y)} = \cos(X \angle Y),
\end{equation}
can be interpreted as a cosine of the angle between $X$ and $Y$. This simple geometric interpretation is handy for understanding some basic properties of the Pearson correlation, including the fact that just like a cosine, $\rho$ is between -1 and 1, and that orthogonality in such context corresponds to the case where {there is no agreement between $X$ and $Y$,} 
that is, $\text{cov}(X, Y) = 0$. The summary measures we develop, introduced in Section~2, are constructed along the same lines, but are based on evaluating the level of agreement between densities (of the biomarkers of diseased and non-diseased subjects) instead of focusing on random variables---as Pearson correlation does. 


A covariate-specific version of our main summary index is here devised, which can be used to assess the discrimination performance of a diagnostic test, conditionally on the value of a covariate. Nonparametric Bayesian estimators for all proposed indexes are developed and the numerical performance of a specific implementation is evaluated in detail through a simulation study. Using Bayesian nonparametric (BNP) inference in a medical diagnostic setting is not unprecedented \citep{erkanli2006, gu2008, inaciodecarvalho2013, rodriguez2014, inaciodecarvalho2015, branscum2015, johnson2015, inaciodecarvalho2016}, {and doing so provides a great deal of flexibility particularly in the dependent case (i.e., a covariate is present).   An additional computational advantage of our covariate-specific summary measure with respect to that of \cite{inaciodecarvalho2013} is that it avoids the need of computing conditional quantiles over a grid of covariates---a task which requires a substantial computational investment.  More importantly, as we elaborate below, our summary measures do not fall into the separation trap depicted in Figure~\ref{motivation}.} 
 
The article is organized as follows. In the next section we introduce the proposed measures along with the corresponding inference tools. In Section~3 we conduct a simulation study. Section~4 offers an illustration of our methods in a prostate cancer diagnosis case study. Proofs are included in the online supplementary materials. 

\section{Geometric measures of diagnostic test accuracy}
\subsection{Angle-based summary measures of diagnostic test accuracy}\label{unconditional}
Our summary measures are built on similar construction principles as Pearson correlation, but instead of looking at the angle between random variables as in \eqref{eq:pearson}, we work directly with the density of the biomarker outcome for diseased and non-diseased subjects, which we denote as $f_{D}$ and $f_{\D}$, respectively. Thus, in place of the covariance inner product we use $\langle f_D, f_{\D}\rangle = \int_{-\infty}^{\infty} f_D(y) f_{\D}(y)\, \dif y$, and in place of the standard deviation (sd) norms, we use $\|f_D\| = \{\int_{-\infty}^{\infty} f_D^2(y)\, \dif y\}^{1/2} < \infty$, and $\|f_{\D}\| = \{\int_{-\infty}^{\infty} f_{\D}^2(y) \, \dif y\}^{1/2} < \infty$. The starting point for the construction of our measure is given by the following standardized inner product: 
\begin{equation}
  \bar{\kappa} = \frac{\langle f_D, f_{\D}\rangle}{\|f_D\| \|f_{\D}\|}.
  \label{kappabar}
\end{equation}

For a medical test with perfect discriminatory ability we would have $\bar{\kappa} = 0$, as $f_D$ would be orthogonal to $f_{\D}$. The higher the value of $\bar{\kappa}$, the lower the discriminatory ability of the corresponding biomarker. Indeed, similar to the Pearson correlation, our measure can be interpreted as an angle between $f_D$ and $f_{\D}$.  However, since $f_D \geqslant 0$ and $f_{\D} \geqslant 0$ it follows that $\langle f_D, f_{\D}\rangle \geqslant 0$, and 
thus the angle between $f_D$ and $f_{\D}$ can only be acute or right, that is $f_D \angle f_{\D}$ is in $[0, \pi/2]$, and thus $\bar{\kappa}$ is in $[0, 1]$. Orthogonality between the biomarker outcome for diseased and non-diseased subjects corresponds to a diagnostic test that perfectly discriminates between diseased and non-diseased subjects. 


{However natural the $\bar{\kappa}$ in \eqref{kappabar} may appear,} in practice one may want to avoid the division by the norms---which in theory could lead to more unstable estimates and constrain us to work with square-integrable densities---while retaining the main ingredients of the construction above. Since $f_D$ and $f_{\D}$ are valid densities, it follows that $\|\sqrt{f_D}\| = \|\sqrt{f_D}\| = 1$, and thus we define our summary measure as
\begin{equation}
    \kappa = \frac{\langle \sqrt{f_D}, \sqrt{f_{\D}} \rangle}
    {\|\sqrt{f_{D}}\| \|\sqrt{f_{\D}}\|} 
    = \langle \sqrt{f_D}, \sqrt{f_{\D}} \rangle 
    = \int_{-\infty}^{+\infty} \sqrt{f_D(y)} \sqrt{f_{\D}(y)} \, \dif y.
  \label{kappa}
\end{equation}
Some comments are in order. Similar to \eqref{kappabar}, orthogonality between the biomarker outcome for diseased and non-diseased subjects corresponds to the case where the diagnostic test is perfect, that is $\kappa = 0$ for a perfect test---which perfectly discriminates diseased subjects from non-diseased subjects---and $\kappa = 1$ for a useless test---for which $f_D = f_{\D}$.

Interestingly, the measure $\kappa$ in \eqref{kappa} is known in mathematical statistics under the name of Hellinger affinity \citep{vaart1998}, but we are unaware of applications of the concept in medical diagnostic statistics. In our context, $\kappa$ can be interpreted as a measure of the level of agreement between the densities of the biomarker outcomes for diseased and non-diseased subjects, or equivalently, as a measure of {the highest possible} diagnostic test accuracy {for a test based on the biomarker}. 

\begin{figure} \centering
  \begin{minipage}{0.275\linewidth}\footnotesize \centering
   $\hspace{-1cm}\kappa = 1$ \\ {$\hspace{-1.4cm} (\AUC = 0.5)$}
  \end{minipage} 
  \begin{minipage}{0.275\linewidth}\footnotesize \centering
   $\hspace{0cm}\kappa = 0.61$ \\ {\hspace{0cm} $(\AUC~= 0.92)$} 
  \end{minipage} 
  \begin{minipage}{0.275\linewidth}\footnotesize \centering
   $\hspace{1.8cm}\kappa = 0.32$ \quad \\ {\hspace{1.5cm} $(\AUC~= 0.98)$} 
  \end{minipage} 
  \begin{minipage}{0.275\linewidth} \vspace{-.5cm}
    \includegraphics[scale = 0.25]{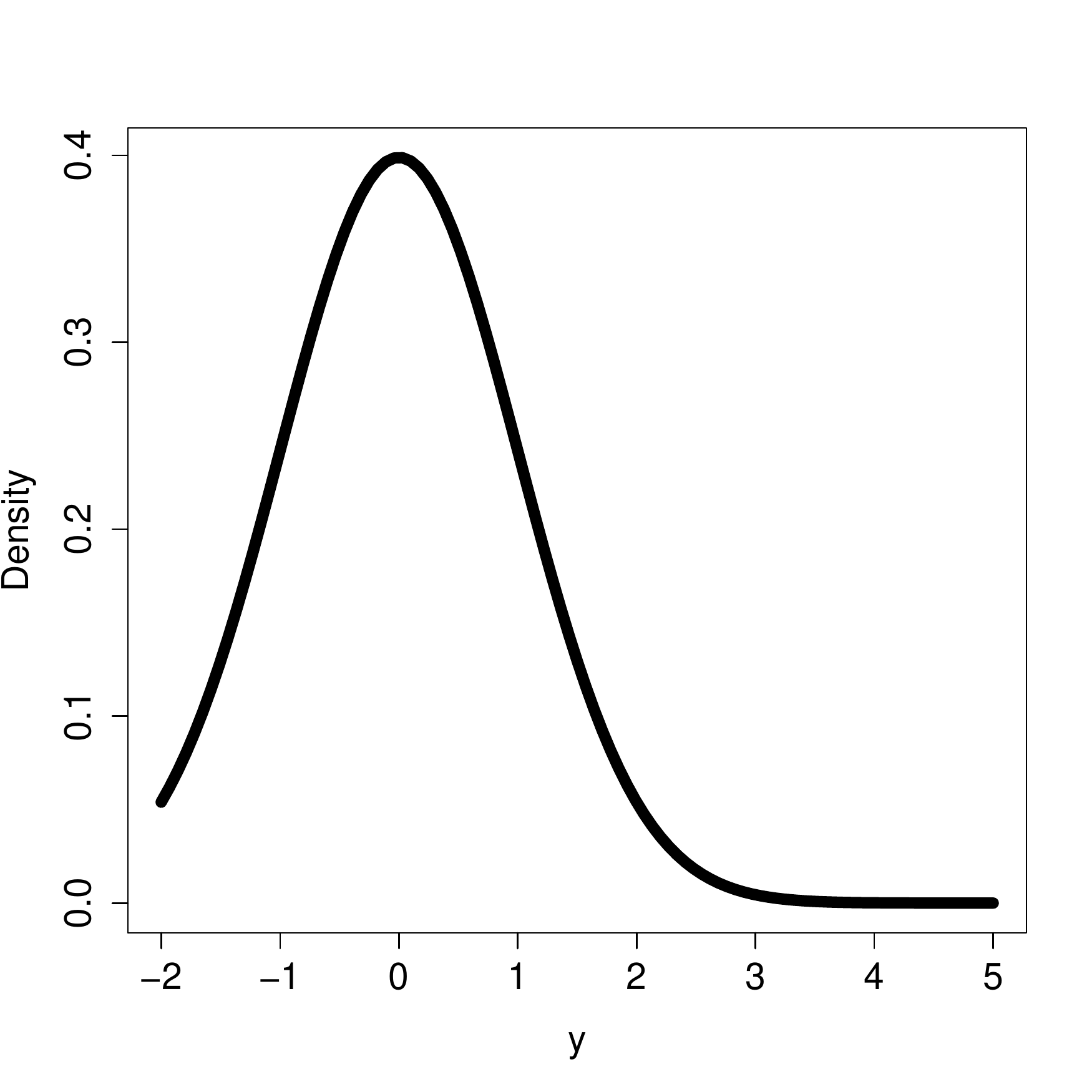}    
  \end{minipage}\hspace{0.7cm}
  \begin{minipage}{0.275\linewidth} \vspace{-.5cm}
    \includegraphics[scale = 0.25]{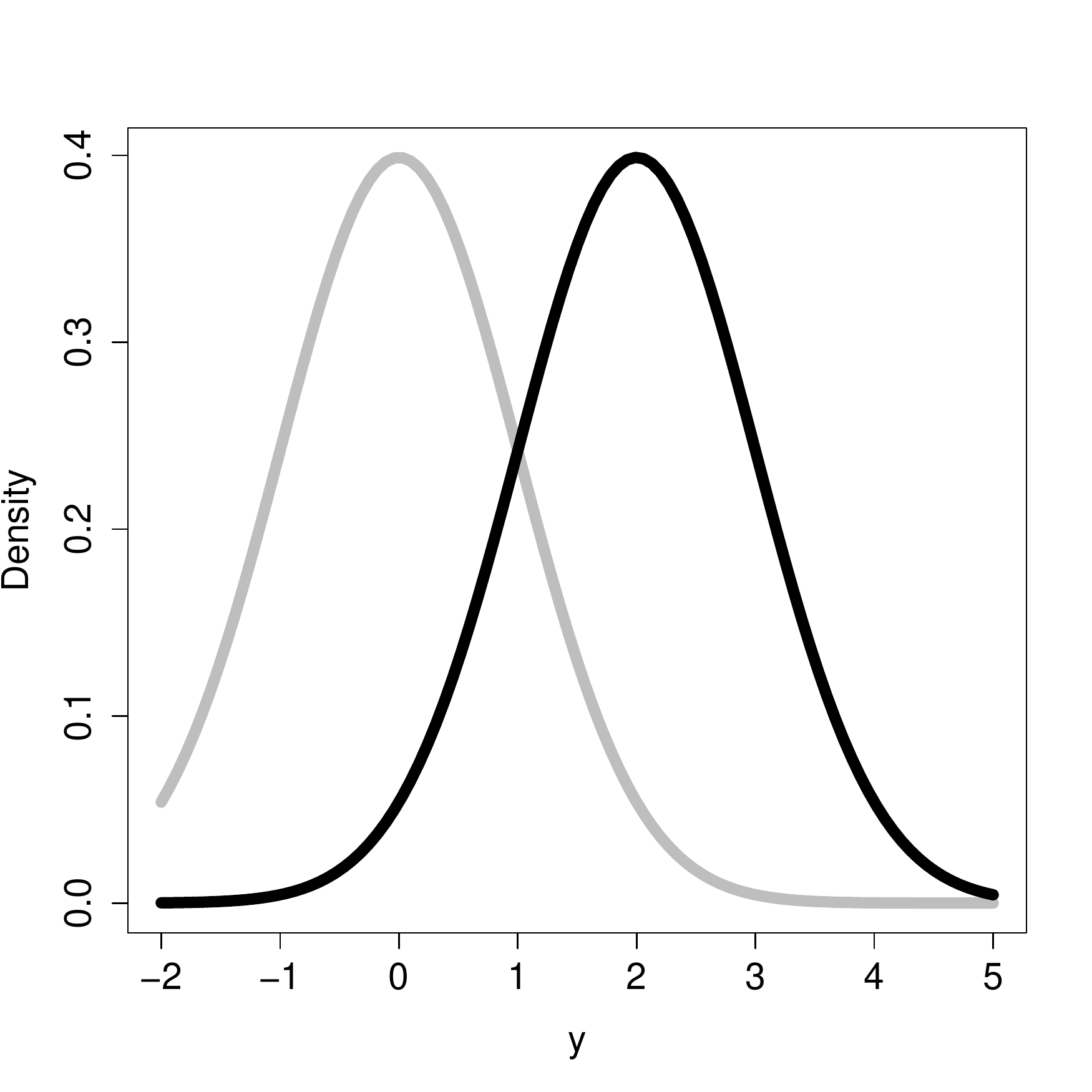}    
  \end{minipage}\hspace{0.7cm}
  \begin{minipage}{0.275\linewidth} \vspace{-.5cm}
    \includegraphics[scale = 0.25]{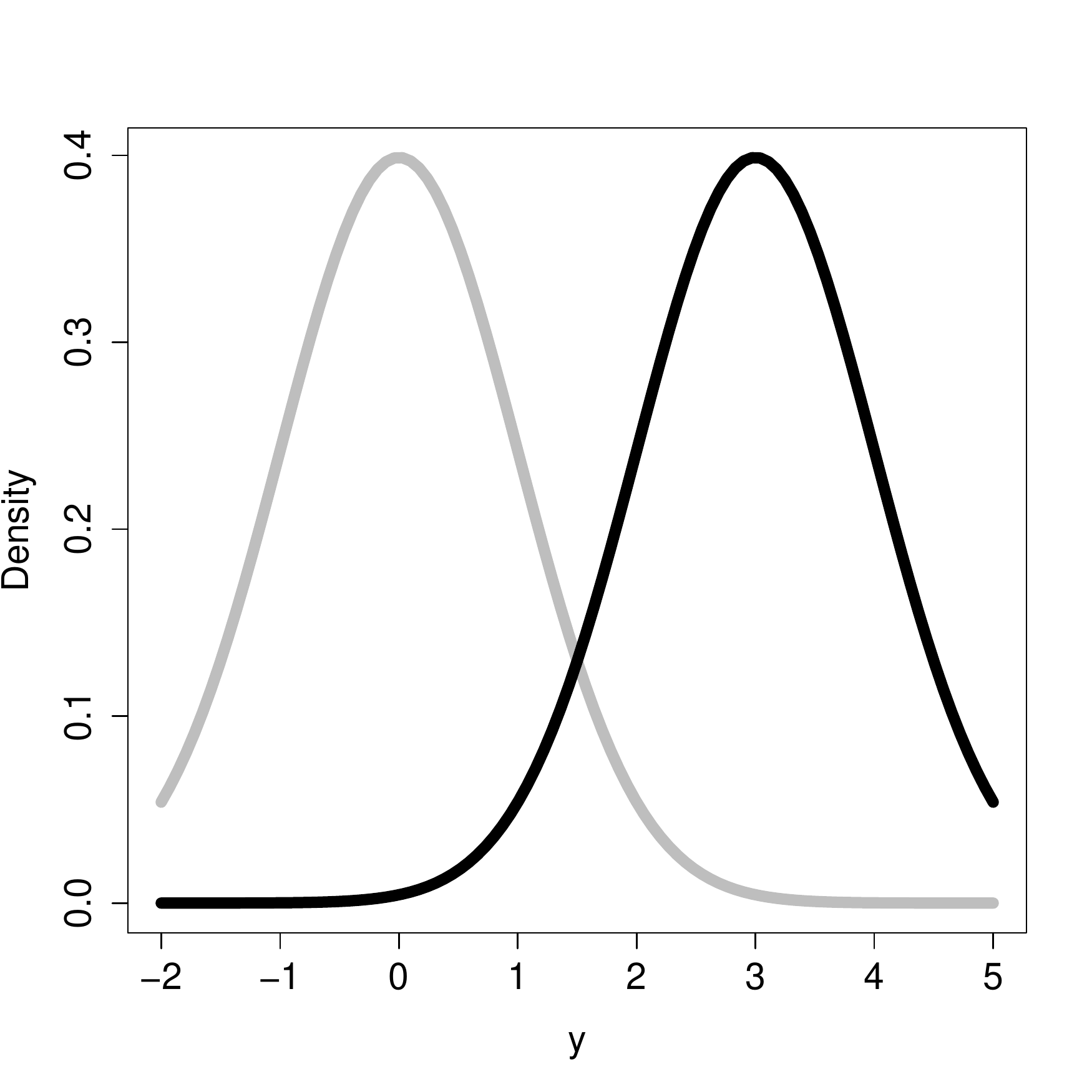}    
  \end{minipage} \\ 
  \begin{minipage}{0.275\linewidth}\footnotesize
    ~~~~~~~~~~~~~~~~~~~(a)
  \end{minipage}\hspace{0.4cm}
  \begin{minipage}{0.275\linewidth}\footnotesize \centering
    ~~~(b)
  \end{minipage}\hspace{0.7cm}
  \begin{minipage}{0.275\linewidth}\footnotesize \centering
    ~~~(c)
  \end{minipage} 
  \caption{\small Affinity for Binormal model from Example~\ref{binormal}; the black  and grey lines respectively denote the densities of the biomarkers of the diseased and non-diseased subjects; the configurations of parameters are as follows: a) $(\mu_D, \sigma_D) = (\mu_{\D}, \sigma_{\D}) = (0, 1)$; b) $(\mu_D, \sigma_D) = (2, 1)$ and $(\mu_{\D}, \sigma_{\D}) = (0, 1)$; c) $(\mu_D, \sigma_D) = (3, 1)$ and $(\mu_{\D}, \sigma_{\D}) = (0, 1)$.}
  \label{comp}
\end{figure}

\begin{example}[Binormal affinity]\normalfont \label{binormal}
  Suppose $f_D(y) = \phi(y \mid \mu_D, \sigma^2_D)$ and $f_{\D}(y) = \phi(y \mid \mu_{\D}, \sigma^2_{\D})$. As stated in Table~\ref{parametric}:
  \begin{equation}
    \kappa = \sqrt{\frac{2 \sigma_D \sigma_{\D}}{\sigma_D^2 + \sigma_{\D}^2}} \exp\bigg\{-\frac{1}{4}
    \frac{(\mu_D - \mu_{\D})^2}{\sigma_D^2 + \sigma_{\D}^2}\bigg\}.
    \label{bincomp}
  \end{equation}
  As expected, for a useless test---that is $\mu_D = \mu_{\D}$ and $\sigma_{D} = \sigma_{\D}$---it holds that $\kappa = 1$. For fixed $\sigma_D > 0$ and $\sigma_{\D} > 0$ it follows that as $\mu_{D} \to \infty$ and $\mu_{\D} \to -\infty$, that is as  populations become more separated, then $\kappa \to 0$. Indeed, as it can be seen from Figure~\ref{comp} the more separated the populations---that is the more orthogonal they are---the closer $\kappa$ gets to zero. Notice also that, in this setting, the larger the $\AUC$ the lower $\kappa$. However, this needs not always be the case, and there are actually situations for which $\AUC$ and $\kappa$ may recommend different decisions, as will be seen in Examples~\ref{trap} and \ref{binormal2}.
\end{example}

\begin{example}[Separation trap]\normalfont \label{trap}
Let's revisit the setting from Figure~\ref{motivation}. The exact setup is \begin{equation*}
  \begin{cases}
    f_{D}(y) = 1/2 \phi_{\textsc{t}}(y \mid -6, -4, -5, 1/3^2) + 
    1/2 \phi_{\textsc{t}}(y \mid 4, 6, 5, 1/3^2), \\
    f_{\D}(y) = \phi_{\textsc{t}}(y \mid -2, 2, 0, 1/4^2).
  \end{cases}
\end{equation*}
Here $\phi_{\textsc{t}}(y \mid a, b, \mu, \sigma^2)$ is the density of a truncated normal with lower bound $a$ and upper bound $b$. In this case it holds that 
\begin{equation*}
  \begin{split}
    \kappa & = \int_{-\infty}^{\infty} \sqrt{f_D(y)} \sqrt{f_{\D}(y)} \, \dif y \\
    & = \int_{-6}^{-4} \sqrt{f_D(y)} \sqrt{f_{\D}(y)} \, \dif y 
    + \int_{-2}^{2} \sqrt{f_D(y)} \sqrt{f_{\D}(y)} \, \dif y  + 
    \int_{4}^{6} \sqrt{f_D(y)} \sqrt{f_{\D}(y)} \, \dif y = 0.
  \end{split}
\end{equation*}
Thus, $\kappa$ claims that both populations are perfectly separated---and so it would not fall into the separation trap. 
\end{example}

Table~\ref{parametric} contains two other examples of affinity for parametric models. For completeness we include derivations of these expressions in the  supplementary materials.

\begin{table}\centering 
\caption{\small Affinity ($\kappa$) for bibeta, biexponential, and binormal models; here, $\lambda_D$ and $\lambda_{\D}$ are the rate parameters of the corresponding exponential distributions}  \footnotesize \vspace{.3cm}
\begin{tabular}{ll} \hline 
Model & Affinity\\
\hline 
Bibeta & $\frac{B((a_D +a_{\D}) / 2, (b_D + b_{\D}) / 2)}{\{B(a_D, b_D)B(a_{\D}, b_{\D})\}^{1/2}}$\\[.3cm] 
Biexponential & $\frac{2(\lambda_D \lambda_{\D})^{1/2}}{\lambda_D + \lambda_{\D}}$\\[.3cm] 
Binormal & $\sqrt{\frac{2 \sigma_D \sigma_{\D}}{\sigma_D^2 + \sigma_{\D}^2}} \exp\bigg\{-\frac{1}{4}
    \frac{(\mu_D - \mu_{\D})^2}{\sigma_D^2 + \sigma_{\D}^2}\bigg\}$\\
\hline
\end{tabular}
\label{parametric}
\end{table}

\subsection{Properties and covariate-specific affinity}
The following proposition documents two elementary properties associated with our measure of diagnostic test accuracy.

\begin{proposition}\label{prop} 
  Affinity, as defined in \eqref{kappa}, obeys the following properties:
  \begin{enumerate}
  \item $\kappa \in [0, 1]$.
  \item $\kappa$ is invariant to monotone increasing data transformations.
  \end{enumerate}
\end{proposition}

{A proof of Proposition~\ref{prop} can be found in the online supplementary materials.} Interestingly, just like affinity, the $\AUC$ is also invariant to monotone increasing data transformations \citep{pepe2003}. Affinity is also invariant to whether we work with a test for which larger values of the biomarker are more indicative of disease, or the other way around; this is an obvious consequence of the fact that $\langle \sqrt{f_D}, \sqrt{f_{\D}} \rangle = \langle \sqrt{f_{\D}}, \sqrt{f_{D}} \rangle$. Thus, for instance, Binormal affinity in \eqref{bincomp} is the same, regardless of whether $\mu_D > \mu_{\D}$ or vice versa. For the lack of better terminology, below we refer to an \textit{upper-tailed diagnostic test} as one for which larger values of the biomarker are more indicative of disease, and to a \textit{lower-tailed diagnostic test} as to one where larger values of the biomarker are less indicative of disease. {Another parallel to the AUC is the fact that $\kappa$ can be regarded as an area under a curve, with the curve of interest being 
\begin{equation*}
  c(y) = \sqrt{f_D(y) f_{\D}(y)}.
\end{equation*}
Another interesting aspect is that $\kappa$ can also be regarded as an average squared likelihood ratio, in the sense that}
\begin{equation*}
  \kappa = \int_{-\infty}^{\infty} \sqrt{\frac{f_D(y)}{f_{\D}(y)}} f_{\D}(y) \, \dif y
    = E_{\D}\bigg(\sqrt{\frac{f_D(Y_{\D})}{f_{\D}(Y_{\D})}} \bigg).
\end{equation*}

If covariates are available the question arises  {of how} to
conduct a covariate-specific analysis for measuring diagnostic test accuracy using affinity. A natural extension of \eqref{kappa} to the conditional setting is  
\begin{equation}
  \kappa(\bmx) = \langle \sqrt{f_{D \mid \bmx}}, \sqrt{f_{\D \mid \bmx}}\rangle = 
  \int_{-\infty}^{+\infty} \sqrt{f_D(y \mid \bmx)} \sqrt{f_{\D}(y \mid \bmx)} \, \dif y, 
\label{kpred}
\end{equation}
where $\bmx \in \mathcal{X} \subseteq \mathbb{R}^p$ is a covariate, $f_{D \mid \bmx} = f_D(\, \cdot \mid \bmx)$, and $f_{\D \mid \bmx} = f_{\D}(\, \cdot \mid \bmx)$. Below we refer to $\kappa(\bmx)$ as the covariate-specific affinity. As with $\kappa$, it holds that $\kappa(\bmx) \in [0, 1]$, and that $\kappa(\bmx)$ is invariant to monotone increasing data transformations.

\begin{figure} \hspace{2cm}
  \begin{minipage}{0.45\linewidth}\hspace{.4cm}
    \includegraphics[scale = 0.3]{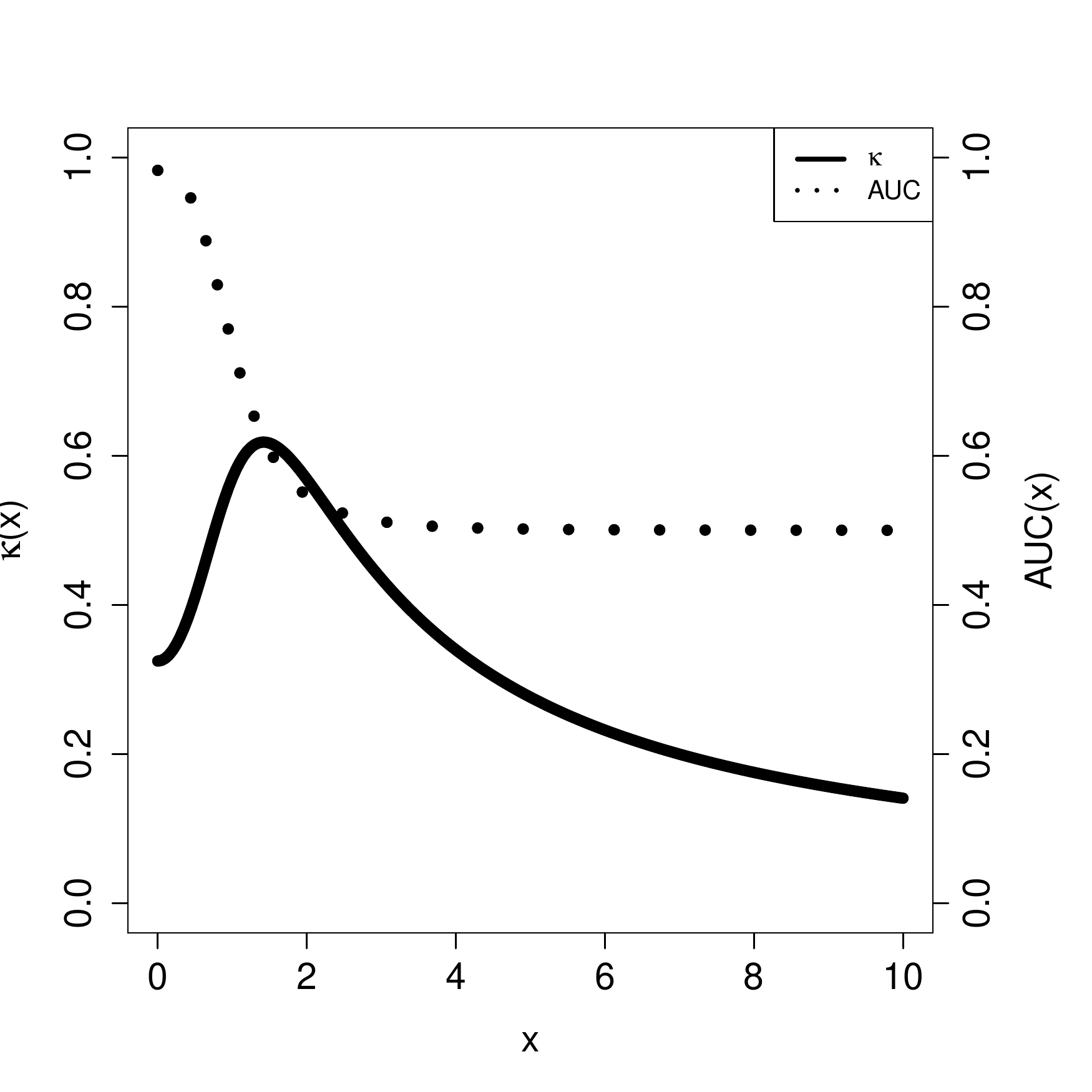}    
  \end{minipage} 
  \begin{minipage}{0.45\linewidth}\hspace{-.4cm}
    \includegraphics[scale = 0.3]{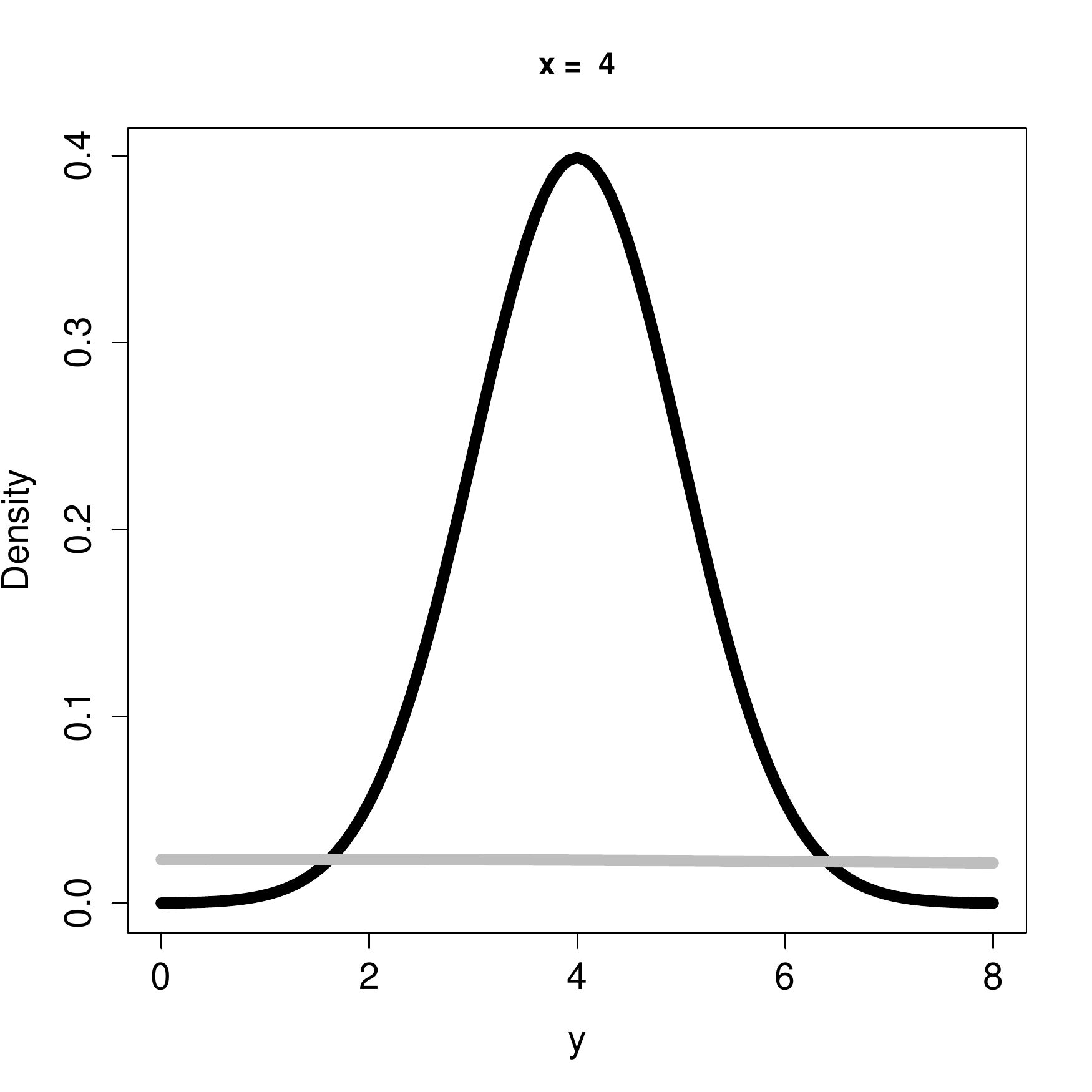}    
  \end{minipage} 
  \begin{minipage}{0.45\linewidth}\centering \footnotesize
    ~~~~~~~~~~~~~~~~~~~~~~(a)
  \end{minipage}
  \begin{minipage}{0.45\linewidth}\centering \footnotesize
    ~~~~~~~~~~~~(b)
  \end{minipage}
  \caption{\small a) Covariate-specific affinity  (solid line) for binormal model from Example~\ref{binormal2}, and corresponding covariate-specific $\AUC$ 
(dotted line); b) Density of the diseased (black line) and non-diseased (grey line) subjects, for $x = 4$.}
    \label{cov}
\end{figure}

\begin{example}[Binormal covariate-specific affinity]\normalfont \label{binormal2}
  Extending Example~\ref{binormal}, suppose that $f_D(y \mid \bmx) = \phi(y \mid \mu_D(\bmx), \sigma^2_D(\bmx))$ and $f_{\D}(y) = \phi(y \mid \mu_{\D}(\bmx), \sigma^2_{\D}(\bmx))$. It then follows that 
  \begin{equation*}
    \kappa(\bmx) = \sqrt{\frac{2 \sigma_D(\bmx) \sigma_{\D}(\bmx)}{\sigma_D^2(\bmx) + \sigma_{\D}^2(\bmx)}} \exp\bigg\{-\frac{1}{4}
    \frac{\{\mu_D(\bmx) - \mu_{\D}(\bmx)\}^2}{\sigma_D^2(\bmx) + \sigma_{\D}^2(\bmx)}\bigg\}.
  \end{equation*}
In particular, for $\mu_D(x) = x$ and $\mu_{\D}(x) = x - 3$, and $\sigma_D(x) = 1$ and $\sigma_{\D}(x) = 1 + x^2$, we obtain the covariate-specific affinity plotted in Figure~\ref{cov}(a). As it can be observed from Figure~\ref{cov}, for values of the predictor between 0 and approximately 1.2, both $\kappa$ and $\AUC$ agree that the quality of the test deteriorates as $x$ increases ($\kappa$ increases and $\AUC$ decreases). {As $x$ increases beyond 1.2, each measure suggests a different conclusion as to how the test accuracy changes with $x$.}  To understand the reason for this, we analyze in further detail the case of $x = 4$, whose corresponding densities are plotted in Figure~\ref{cov}(b). In the case $x = 4$ we have an $\AUC = 0.504$ whereas $\kappa = 0.34$. Thus, on the one hand the $\AUC = 0.504$ suggests that the test is quite poor, whereas the value of $\kappa = 0.34$ suggests that it could be satisfactory, though far from excellent. The intuition underlying this lack of agreement is as follows: $\kappa$ is taking into account that around 95\% of the mass for the test values for diseased subjects will be on the $[0, 8]$ interval, whereas around 95\% of the mass for the test values of non-diseased subjects will be on the $[-30, 38]$ interval. 
\end{example}

\subsection{Nonparametric Bayesian inference for affinity and covariate-specific affinity}\label{NonparInf}
In this section we discuss Bayesian nonparametric estimators for affinity, as defined in \eqref{kappa}, and covariate-specific affinity, as defined in \eqref{kpred}. {Let $\{Y_{D, i}\}_{i = 1}^{n_D}$ and $\{Y_{\D, i}\}_{i = 1}^{n_{\D}}$ be random samples from $F_{D}$ and $F_{\D}$}. We propose to estimate $\kappa$ in \eqref{kappa} by modeling each conditional density $f_D$ and $f_{\D}$ as an infinite mixture model of the type
\begin{equation}
  f(y) = \int_{\Theta} K(y \mid \thetab) \, G(\dif \thetab), 
  \label{mixgen}
\end{equation}
where $K$ is a kernel and $G$ is a random mixing measure. The corresponding induced prior is 
\begin{equation}
  \kappa = \int_{-\infty}^{+\infty} \bigg\{\int_{\Theta} K(y \mid \thetab) \, G_D(\dif \thetab)\bigg\}^{1/2} \bigg\{\int_{\Theta} K(y \mid \thetab) \, G_{\D}(\dif \thetab)\bigg\}^{1/2} \, \dif y.
\label{plugin}
\end{equation}
A natural approach is to consider each $G$ as a Dirichlet process \citep{ferguson1973}, and to rely on normal kernels, in which case \eqref{mixgen} becomes a so-called Dirichlet process mixture of normal kernels,  
\begin{equation}
  f(y) = \int_{\mathbb{R} \times (0, \infty)} \phi(y\mid \mu, \sigma) \,
  G(\dif \mu, \dif \sigma) = \sum_{h=1}^\infty \pi_h \phi(y\mid \mu_h, \sigma_h), \quad G \sim \text{DP}(\alpha, G_0).
  \label{mixDPM}
\end{equation}
Here $\alpha > 0$ is the so-called precision parameter, $G_0$ is the centering distribution function, or baseline measure, and we use the notation $G \sim \text{DP}(\alpha, G_0)$ to represent that $G$ follows a Dirichlet process (DP). 
A celebrated representation of the DP is the so-called stick-breaking construction \citep{sethuraman1994}. According to this representation a random distribution function $G$ follows a DP if it admits a representation of the type
\begin{equation}
  G = \sum_{h=1}^{\infty}\pi_h \delta_{\thetab_h} , \quad \thetab_h \overset{\text{iid}}{\sim} G_0,  
  \label{stick_break}
\end{equation}
where $\pi_1 = V_1$, and 
$\pi_h = V_h\prod_{k < h}(1 - V_k)$, with $V_h \iid \text{Beta}(1,\alpha)$, for $h = 2, \dots$. The $\thetab_h$ are known as atoms, the $\pi_h$ as mixing weights, and the $V_h$ are the so-called stick-breaking weights. 

{For regression data, $\{(\bmx_i, Y_{D, i})\}_{i = 1}^{n_D}$ and $\{(\bmx_i, Y_{\D, i})\}_{i = 1}^{n_{\D}}$, we} propose to estimate $\kappa(\bmx)$ in \eqref{kappa} by modeling each density $f_D$ and $f_{\D}$ as an infinite mixture model of regressions 
\begin{equation}
  f(y \mid \bmx) = \int_{\Theta} K(y \mid \thetab) \, G_{\bmx}(\dif \thetab), 
  \label{mixgenreg}
\end{equation}
where $K$ is a kernel and $G_{\bmx}$ is a covariate-specific random mixing measure. The corresponding induced prior is 
\begin{equation}
  \kappa(\bmx) = \int_{-\infty}^{+\infty} \bigg\{\int_{\Theta} K(y \mid \thetab) \, G_{D, \bmx}(\dif \thetab)\bigg\}^{1/2} \bigg\{\int_{\Theta} K(y \mid \thetab) \, G_{\D, \bmx}(\dif \thetab)\bigg\}^{1/2} \, \dif y.
\label{plugin}
\end{equation}
A natural approach is to consider each $G_{\bmx}$ as a dependent Dirichlet process (DDP) \citep{maceachern2000}, and to rely on normal kernels in which case \eqref{mixgenreg} becomes an infinite mixture of regression models,  
\begin{equation}
  f(y \mid \bmx) = \int_{\mathbb{R} \times (0, \infty)} \phi(y\mid \mu, \sigma) \,
  G_{\bmx}(\dif \mu, \dif \sigma).
  \label{mixDP2}
\end{equation}
Because of the support properties in Theorem~4 of \cite{barrientos2012}, we consider a `single-weights' DDP \citep{deiorio2004, deiorio2009}
\begin{equation}\label{ddp}
G_{\bmx} = \sum_{h=1}^{\infty}\pi_{h}\delta_{\thetab_{\bmx, h}}.
\end{equation}
The random support locations $\thetab_{\bmx,h}$ are, for $h = 1,2,\ldots$ independent and identically distributed realizations from a stochastic process over the covariate space $\mathcal{X}$ and the weights $\{\pi_h\}_{h = 1}^{\infty}$ match those from a standard DP; in this specific version of \eqref{ddp} we obtain
\begin{equation}
  f(y \mid \bmx) = \sum_{h=1}^\infty \pi_h \phi(y\mid \mu_h(\bmx), \sigma_h). 
  \label{cond_ddp}
\end{equation}
To achieve a reasonable tradeoff between flexibility and parsimony, in practice we choose to model $\mu(\bmx)$ as a linear model, that is, $\mu_h(\bmx) = \bmx^{\T} \betab_h$, where $\bmx^{\T}$ corresponds {to the cubic B-spline basis evaluated at the predictor.   Finally, {to facilitate prior specification we suggest} \textit{standardizing} the {biomarkers} (i.e., $Z_{Di} =(Y_{Di} - \bar{Y}_{D})/s_{D}$ and $z_{\bar{D}j}=(Y_{\bar{D}j}- \bar{Y}_{\bar{D}})/s_{\bar{D}}$) {and rescaling the covariate (i.e., $\min\{ x_{\bar{D}}, x_{D}\}=-1$ and $\max\{x_{\bar{D}}, x_{D}\}=1$)}.  Having estimated the densities on the standardized data, the location-scale adjustment may be applied to easily convert to densities for the untransformed data.

{We now present a specific embodiment of our model.} Let $\bmx_{\bar{D}i}^{\T}$ represent the cubic B-spline representation of $x_{\bar{D}i}$, with $x_{\bar{D}i}$ having been rescaled to lie in $[-1,1]$. The assumptions for the non-diseased population in the conditional case are that 
\begin{align}\label{likelihood}
f_{\bar{D}}(Z_{\bar{D}i} \mid x_{\bar{D}i}) & = 
  \int {\phi(Z_{\bar{D}i} \mid \bmx_{\bar{D}i}^{\T} \betab, \sigma^2)} \,\dif G_{\bar{D}}(\betab, \sigma^2)  \\
G_{\bar{D}}( \betab, \sigma^2) \mid G_{\bar{D}0}(\betab, \sigma^2) & \sim \DP(1, G_{\bar{D}0}( \betab, \sigma^2)) \\
G_{\bar{D}0}( \betab, \sigma^2) & \equiv \mbox{N}(\betab_{\bar{D}0}, \Sigma_{\bar{D}0}) \times \mbox{IG}(\mbox{shape}=1, \mbox{rate}=50)\\
\beta_{\bar{D}0} & \sim \mbox{N}(0, I) \\
\Sigma_{\bar{D}0} & \sim \mbox{IWish}(\nu=1, S= I), 
\end{align}
where IG and IWish respectively denote the inverse Gamma and inverse Wishart distributions. Two aspects of this specification are particularly noteworthy. First, it is assumed that the number and locations of all knots are known, although this could be relaxed. Second, the prior on the within-cluster variance (i.e., $\sigma^2$) was chosen to favor variances much less than one.  The justification for this is immediate when recognizing that the likelihood is on standardized data with a marginal sample variance of one; the within-cluster variance ought to be substantially smaller than the marginal variance. The assumptions are analogous for the diseased population; the only difference is the substitution of $D$ for $\bar{D}$.  To apply the model specification without conditioning on any covariate, we can simply substitute 
{$\bmx_i^{\T}=1$.}

\subsection{Theoretical properties on induced priors}\label{theory}
This section includes theoretical properties on the induced priors for the summary measures introduced in Section~\ref{unconditional}. {Although in practice we model the densities from which $\kappa$ is estimated with a Dirichlet process mixture, as in {\eqref{mixDPM}} and \eqref{cond_ddp}, below we document theoretical results which apply more generally to \eqref{mixgen} and \eqref{mixgenreg} and only require that the mixing distribution has a full weak support, which includes the Dirichlet process as a particular case. In what follows, we assume the same setting as in \cite{lijoi2004}, namely: 
\begin{enumerate}
\item[A$_1$)] The random mixing distribution(s) has (have) full weak support. 
\item[A$_2$)] $\int_{-\infty}^{+\infty} K(y \mid \thetab) \, \dif y = 1$, for $\thetab \in \Theta$.
\item[A$_3$)] $\thetab \mapsto K(y \mid \thetab)$ is bounded, continuous, and $\mathbb{B}_{\Theta}$-measurable for $y \in \mathbb{R}$.
\item[A$_4$)] The family of mappings $\{\thetab \mapsto K(y \mid \thetab): y \in C\}$, is uniformly equicontinuous, for every compact $C \subset \mathbb{R}$.
\end{enumerate}
Here, A$_1$ is a condition on the support of the mixing, whereas A$_2$--A$_4$ are regularity conditions on the kernel. Under these conditions, it can be shown that $f(y)$ in \eqref{mixgen} has full Hellinger support \citep{lijoi2004}. As a consequence, the following result holds. 

\begin{theorem}\label{thm1} \normalfont
Suppose A$_1$--A$_4$ and let $(\Omega, \mathcal{A}, P)$ be the probability space associated with the infinite mixture model in \eqref{mixgen}, which induces $\kappa = \int \sqrt{f_D(y)} \sqrt{f_{\D}(y)} \, \dif y$. Let $\kappa^\omega$ be a realization of the $\kappa$ index under \eqref{mixgen}. Then, for every $\varepsilon > 0$, it holds that $P\{\omega \in \Omega: |\kappa^\omega - \kappa| < \varepsilon\} > 0$.
\end{theorem}

Under the same conditions as above, it can be shown that $f(y \mid \bmx)$ in \eqref{mixgenreg} has full Hellinger support \citep{barrientos2012}. Thus, the following analogous result to Theorem~\ref{thm1} holds for the covariate-specific version of our summary measure as defined in \eqref{kpred}. 

\begin{theorem}\label{thm2} \normalfont
Suppose A$_1$--A$_4$ and let $(\Omega, \mathcal{A}, P)$ be the probability space associated with the infinite mixture of regression models in \eqref{mixgenreg}, which induces $\kappa(\bmx) = \int \sqrt{f_D(y \mid \bmx)} \sqrt{f_{\D}(y \mid \bmx)} \, \dif y$. Let $\kappa^\omega(\bmx)$ be a trajectory of covariate-specific affinity $\kappa(x)$ under \eqref{mixgenreg}. Then, for $\bmx_1,\ldots,\bmx_n \in \mathcal{X}$, for every positive integer $n$ and $\varepsilon>0$, it holds that $P\{\omega \in \Omega: |\kappa^\omega(\bmx_i) - \kappa(\bmx_i)| < \varepsilon,~i=1,\ldots,n\} > 0$.
\end{theorem}

Proofs of Theorems~\ref{thm1} and~\ref{thm2} can be found in the online supplementary materials.
\begin{figure} \centering
    \includegraphics[scale = 0.77]{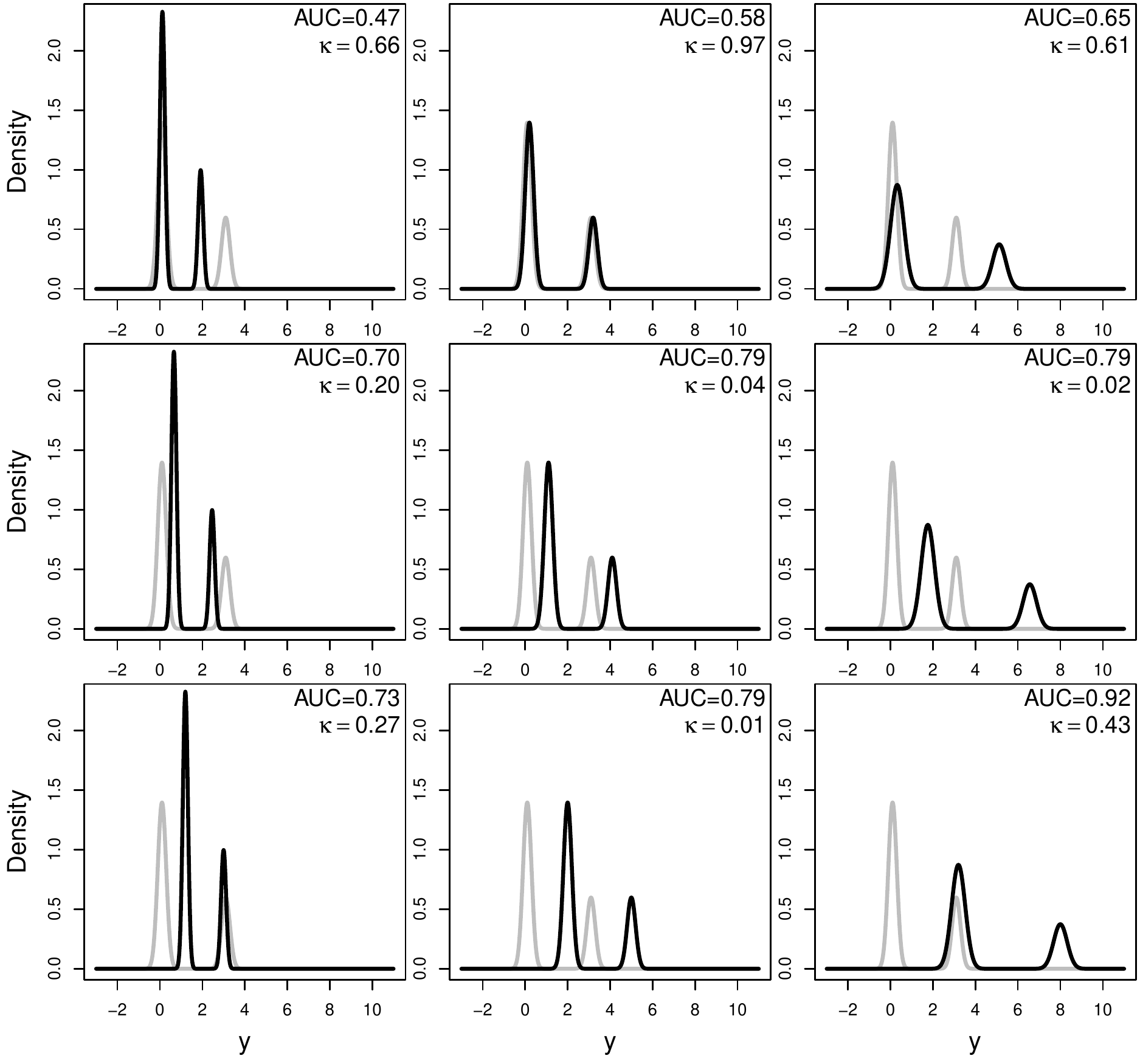}
   \caption[Distributions from Unconditional Setting 2]{
     Densities for the second unconditional simulation study 
     setting in Table~\ref{simset}; the black and grey lines respectively denote the densities of the biomarkers of the diseased and non-diseased subjects.}
    \label{simsetu2}
\end{figure}

\section{Simulation study}

\subsection{Data generating processes}\label{dgp}

The simulation settings are summarized in Table~\ref{simset}. The simulation employed pairs of biomarker distributions that were either conditional on a single uniformly distributed covariate or were unconditional.  In the unconditional settings, each distribution was either normal or a mixture of normals, and the means and standard deviations were systematically altered so that a range of $\kappa$ and $\AUC$ values were considered. In the conditional settings, the conditional distributions were likewise normal or a mixture of normals, and the covariate's effect on the mean and standard deviation were altered to have varying levels of complexity. In terms of the conditional setting, we consider the same scenarios as in \cite{inaciodecarvalho2013}.

\begin{table}
\caption{\small Simulation study settings}  
\begin{center}
\resizebox{\textwidth}{!}{%
\begin{tabular}{lllc}
\hline 
{Scenario} &  {Non-Diseased} ($f_{\bar{D}}$) & {Diseased} ($f_{D}$) & Notes$^\dagger$ \\ \midrule
Unconditional \#1 & $\phi(.4, .8)$ & $\phi(\mu_D, \sigma_D)$ & 1) \\ [0.25 cm]
Unconditional \#2 & $.7\phi(.1,.2) + .3\phi(3.1, .2)$ & $.7\phi(\mu_{1D},  \sigma_D) + .3\phi(\mu_{2D}, \sigma_D)$ & 2)  \\ [0.25 cm]
Conditional \#1 & $\phi(.5+x_{\bar{D}}, 1.5)$ & $\phi(2+4x_D, 2)$ & 3)  \\ [0.25 cm]
Conditional \#2 & $\phi(\sin(\pi (x_{\bar{D}}+1)), .5)$ & $\phi( .5+x_D^2, 1)$   & 3)  \\ [0.25 cm]
\multirow{2}{*}{Conditional \#3} & $\phi(\sin(\pi x_{\bar{D}}), \sqrt{.2+.5\exp(x_{\bar{D}})})$ & $(1+\exp(-x))^{-1}\phi( x_D, .5) + (1+\exp(x))^{-1}\phi( x_D^3, 1)$   &   3) \\ [0.1 cm]
\bottomrule
\end{tabular}}
{\footnotesize $^\dagger$Each unconditional setting has nine distinct pairs of means and variances where for 1): $\mu_D$ in $\{.8, 1.6, 3.2\}$ and $\sigma_D$ in $\{.8, 1.2, 1.6  \}$ and 2): $\sigma_D =  .2c$ and $(\mu_{1D}, \mu_{2D})$ in $\{(.2c, 3.2c), (1.1c, 4.1c), (2c, 5c) \}$, with $c$ in $\{.6, 1, 1.6\}$. For the conditional setting we have 3): $x_{\bar{D}}, x_{D} \iid \mbox{Unif}(-1,1)$.}
\end{center}
\label{simset}
\end{table}

\begin{figure} \hspace{2cm}
  \begin{minipage}{0.45\linewidth} \hspace{-.5cm} 
    \includegraphics[scale = 0.4]{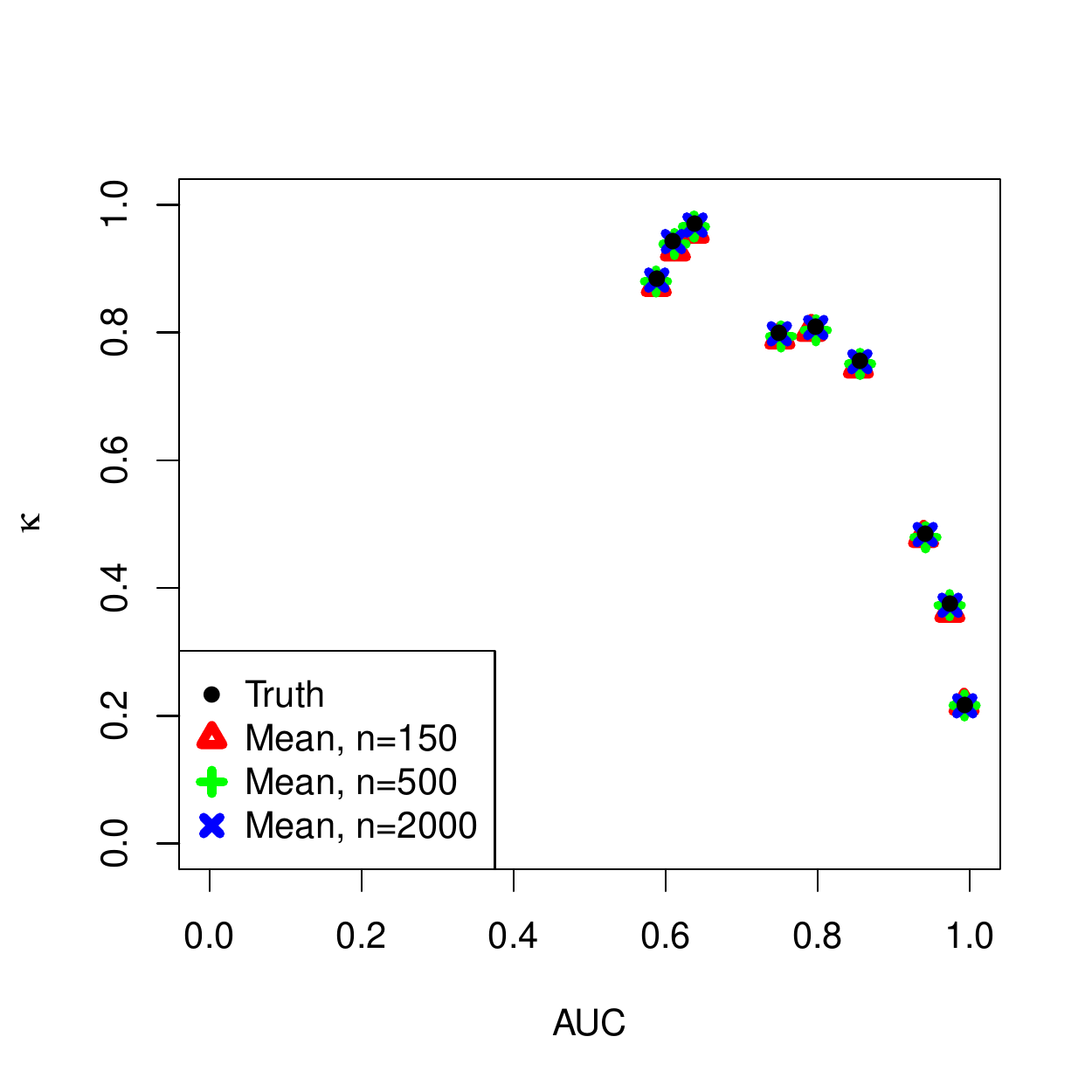}    
  \end{minipage}\hspace{0.7cm}
  \begin{minipage}{0.45\linewidth} \hspace{-.5cm} 
    \includegraphics[scale = 0.4]{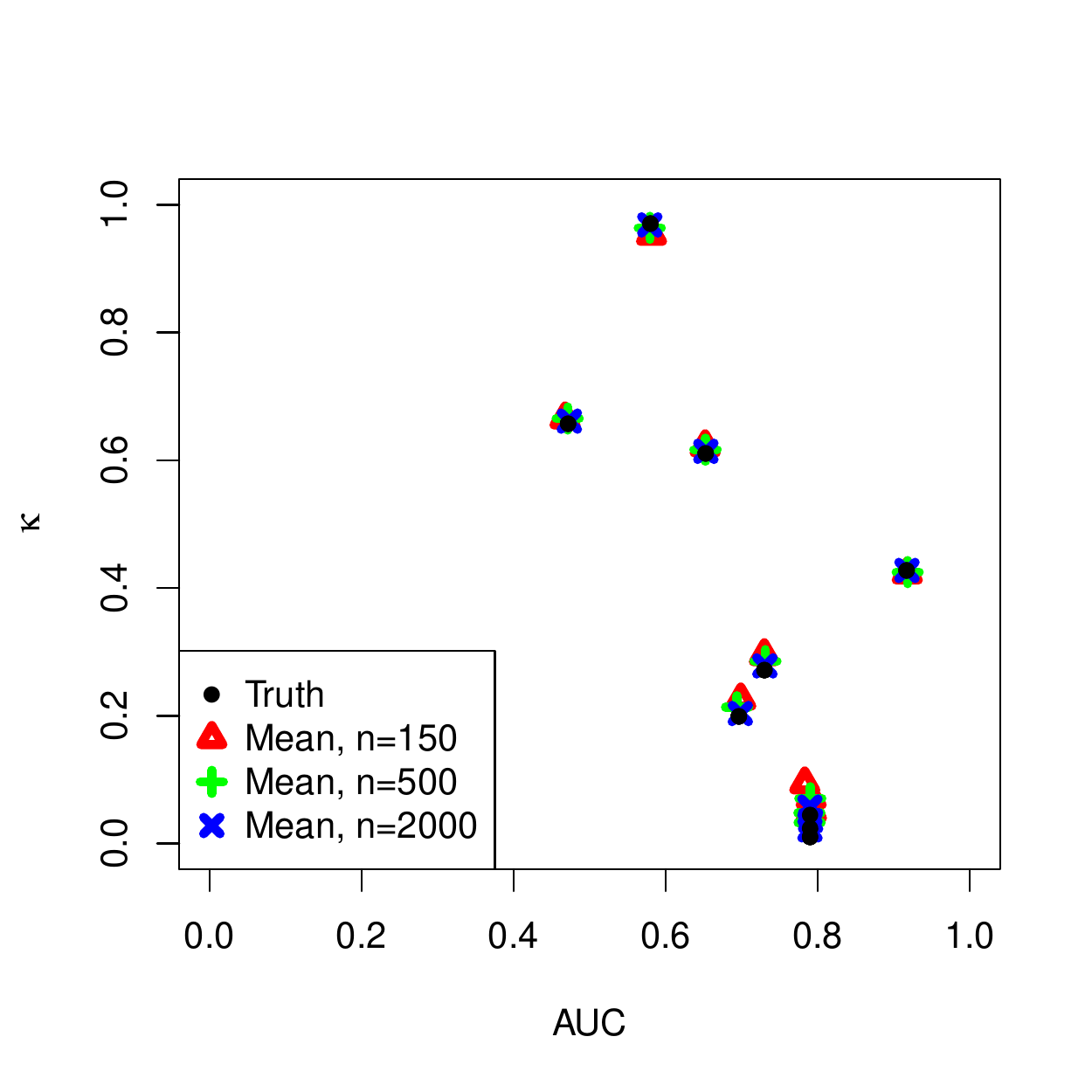}    
  \end{minipage} \\
  \begin{minipage}{0.45\linewidth} \footnotesize
    \hspace{4.2cm}(a)
  \end{minipage}\hspace{0.7cm}
  \begin{minipage}{0.45\linewidth} \footnotesize
    \hspace{4.2cm}(b)
  \end{minipage} 
  \caption{\small $\AUC$ and $\kappa$ estimates (average across 100 simulations) along with true values in the unconditional scenarios of the simulation study (Table~\ref{simset}): a) the first unconditional setting (normals); b) the second unconditional setting (mixtures of normals).}
  \label{fig:u1u2results}
\end{figure}

Figure~\ref{simsetu2} depicts the density pairs from the second unconditional {setting}; the plots for the {remaining} scenarios are included in the supplementary materials. Of particular note is the pattern of possibilities for $\kappa$ and $\AUC$ when the biomarker densities are mixtures of normals. In particular, the middle plot in Figure~\ref{simsetu2} displays a situation where $\kappa$ is particularly adept at identifying the distinctiveness of the diseased and non-diseased populations {as can be seen by the very small $\kappa$ value}. However, our convention that $\AUC$ be computed assuming the diagnostic test will be one-sided forces the $\AUC$ to be lower  than might be expected given the distinctiveness of the populations. While it is certainly possible to entertain more flexible regions at which the diagnostic test would be considered to have a positive result, this would require another nontrivial step before $\AUC$ could even be calculated, whereas such a step is not needed to calculate $\kappa$. 

\subsection{Monte Carlo simulation study}\label{MonteCarlo}
For each setting in Table~\ref{simset}, we generated 100 data sets from $f_D$ and from $f_{\bar{D}}$. The sample sizes were varied at $n_D=n_{\bar{D}}=150$, 500, or 2000 to provide some sense of how reliably $\kappa$ and $\AUC$ were estimated in moderate to large samples. {In implementing the model, detailed in \eqref{likelihood}, no additional knots for the cubic B-splines were included; this lets us ascertain the covariate-dependent model's flexibility in the absence of extra knots.} {Additionally, because the covariate values were simulated from the Unif(-1,1) distribution, we did not rescale the covariate prior to computing the B-spline representation.}  {For each synthetic data set $\kappa$ was estimated by collecting 300 MCMC iterates after a burn in of 2000 and thinning of 40.  Algorithm~8 of \cite{neal2000} was employed to  collect the iterates.} 

{The results from the unconditional settings are summarized in} Figure~\ref{fig:u1u2results}, which depicts the Monte Carlo average (across 100 simulations) of the estimated values for $\kappa$ and $\AUC$, along with the actual values. In part a), which was characterized by each population having a normal distribution, $\kappa$ and $\AUC$ were both estimated with little bias. Not surprisingly, the bias is reduced by having larger sample sizes. In part b), which was characterized by each population having a mixture of normals distribution, the same pattern was exhibited.

\begin{figure}[H]\centering 
  \includegraphics[scale = 0.525]{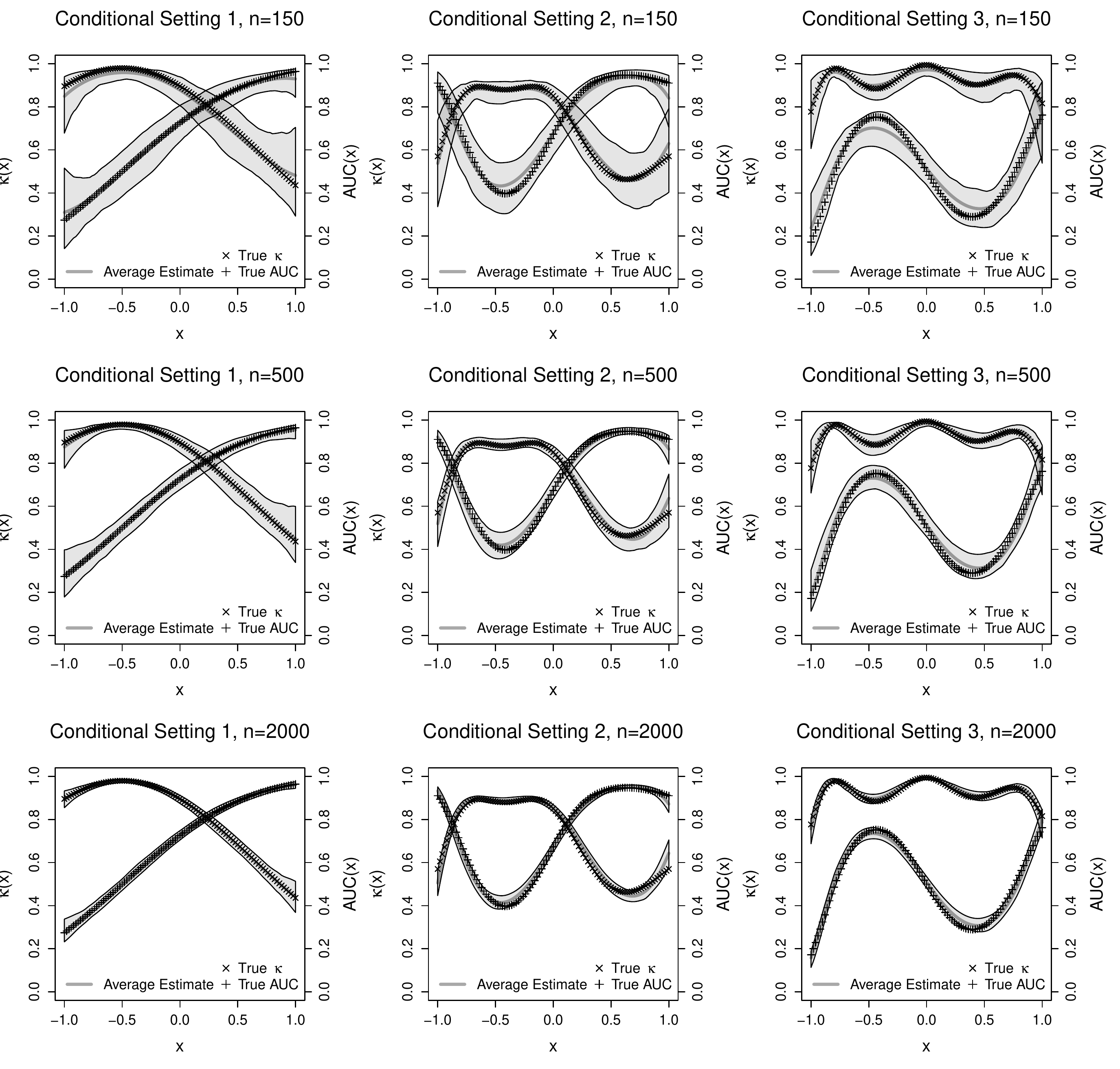} 
  \caption{\small Estimated covariate-specific affinity, $\kappa(x)$, and covariate-specific AUC, $\AUC$$(x)$, across the 100 simulated data sets for the conditional settings described in Table~\ref{simset}. The bands represent the pointwise empirical 2.5th and 97.5th percentiles of the 100 point estimates, while the dark grey lines represent the average of the 100 estimates.}
  \label{fig:condplot}
\end{figure}

The results from the conditional settings are summarized in Figure~\ref{fig:condplot}. For each of the 100 simulated data sets, the conditional means for $\kappa$ and $\AUC$ were estimated at values of $x$ ranging from -1 to 1. The pointwise averages of the 100 estimated means are plotted in this figure, as well as the 2.5th and 97.5th empirical percentiles of these estimated means. This gives some sense for how variable the estimates are (primarily attributable to differences between the 100 simulated data sets). 
{Point estimates of $\AUC(x)$ and $\kappa(x)$ were quite successful in estimating the corresponding true values.  Predictably, the estimates exhibited less variability as more data were available. Recall that a strength of $\kappa(x)$ is that it is not susceptible to the separation trap, nor does it require us to distinguish between upper- and lower-tailed diagnostic tests. This distinction for $\AUC(x)$ explains why the AUC is sometimes estimated to be well below 0.5. Given these advantages of $\kappa(x)$ over $\AUC(x)$, it is even more notable that $\kappa(x)$ can be reliably estimated. An important collateral suggestion of the simulation is that the model is quite flexible even if the cubic B-spline basis does not include additional knots, though of course knots may be added if desired.}

\section{Revisiting a  prostate cancer diagnosis study}
{We now turn our attention to an application that has been regularly employed} {to demonstrate diagnostic test accuracy that is covariate-dependent.}

\begin{figure}[H]
\begin{center}
\includegraphics[scale=0.5]{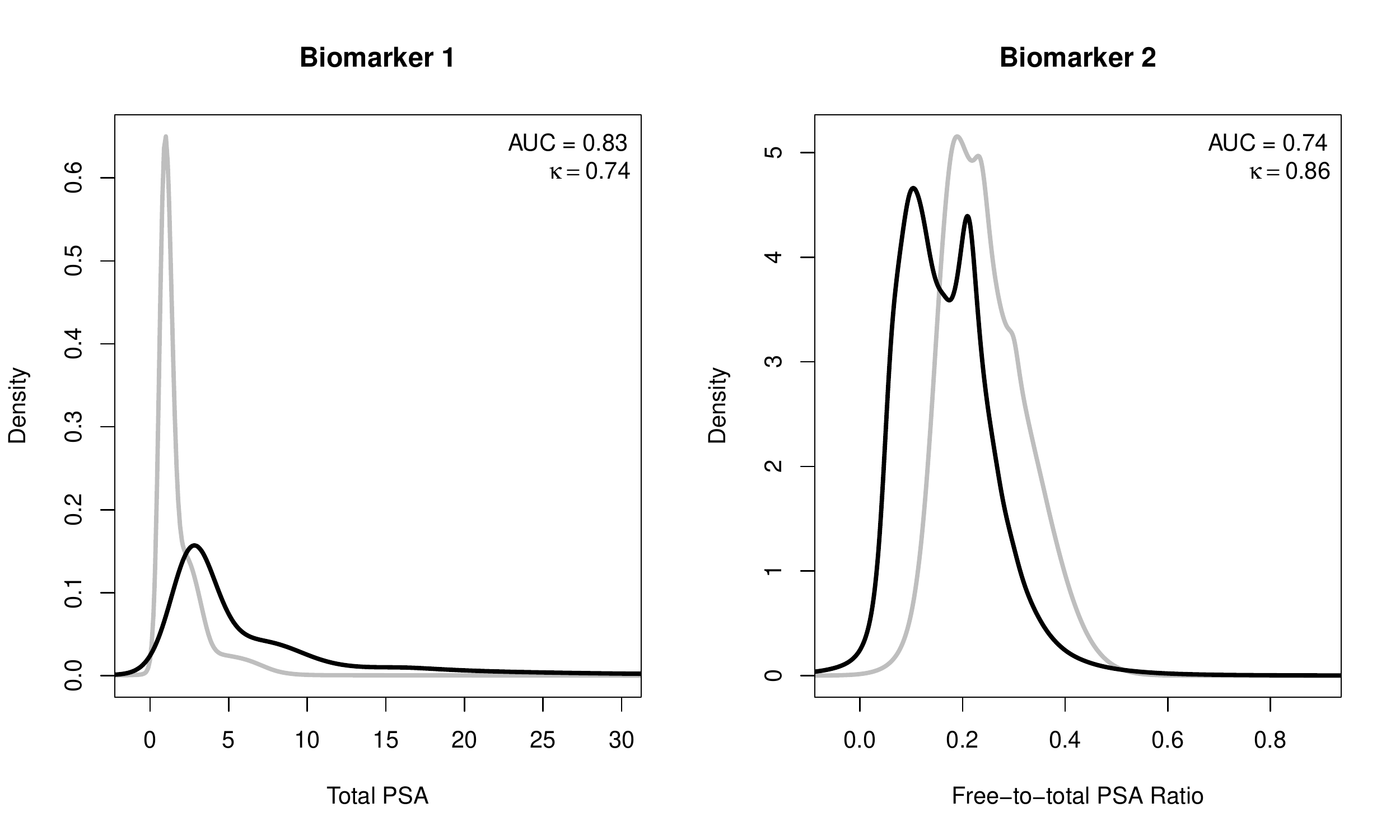}\\ \vspace{-4.2cm}
\includegraphics[scale=0.7]{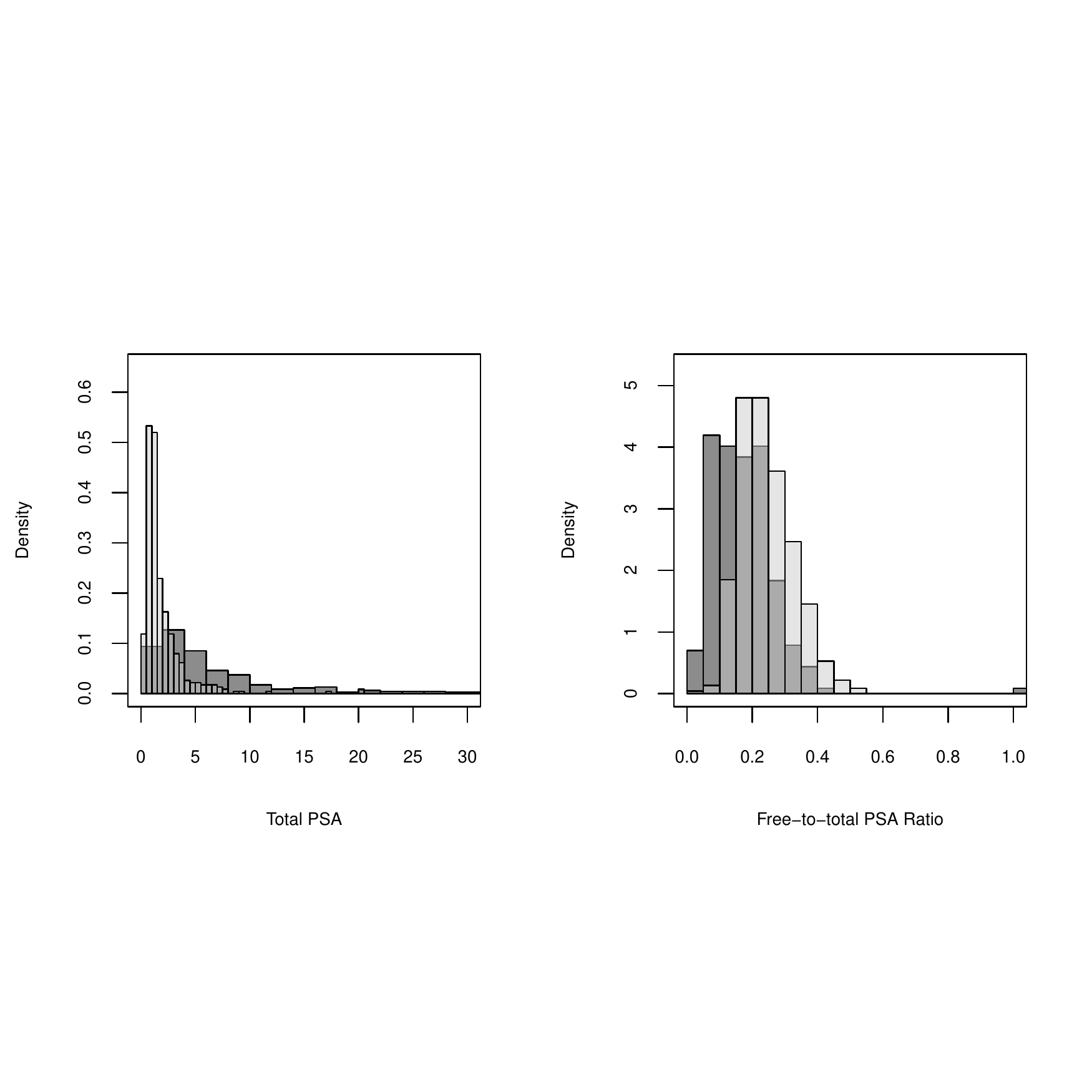} \\  \vspace{-3.3cm}
\caption{\small Top: DPM-based estimated densities along with AUC and $\kappa$ values when age is not considered.  The black and grey lines respectively denote the densities of the biomarkers of the diseased and non-diseased subjects. Bottom: Overlapping histograms.}
\label{Figure:UnconditionalCancer}
\end{center}
\end{figure}

\subsection{Study data and preliminary considerations}

The data were gathered from the Beta-Carotene and Retinol Efficacy Trial (CARET)---a lung cancer prevention trial, conducted at the Fred Hutchinson Cancer Research Center. During this study longitudinal measurements of two Prostate Specific Antigen (PSA)-based biomarkers were collected for 71 prostate cancer cases and 70 controls. The biomarker measurements were taken on males between 46 and 80 years old. The number of repeated measures per subject ranged from one to nine, with $n = 683$ total observations. Further details on this study can be found, for instance, in \cite{etzioni1999} and \cite{pepe2003}. To make our inferences directly comparable with those of \cite{rodriguez2014}---who consider a Gaussian process prior-based model for $\AUC(x)$---we follow the latter authors and ignore the longitudinal nature of the data; however, for reference, we also include in the supplementary materials the results from restricting analysis to each subject's last available observation. A test based on total PSA concentration (Biomarker~1, ng/ml) was assumed to have a positive test result if the measurement was sufficiently large. Conversely, a test based on the free-to-total PSA ratio (Biomarker~2, f/t) was assumed to have a positive test result if the measurement was sufficiently small. 

The direction of the tendency is of no consequence in estimating $\kappa$. In estimating AUC, however, we must consider the direction of the diagnostic test, that is whether larger values of the biomarker are more indicative of disease or the other way around.  A main goal below will be on illustrating how the proposed methods can be used to assess which biomarker might screen better for prostate cancer.  

\subsection{PSA-based analysis}\label{uncondData}


{We first fit the unconditional model (i.e., sans covariate so that 
{$\bmx_i^{\T}=1$}) detailed in \eqref{likelihood} by collecting 1\,800 MCMC iterates after a burn in of 20\,000 and thinning of 100.}  To visualize differences between the biomarkers we provide Figure~\ref{Figure:UnconditionalCancer}. For each biomarker, the estimated density among cases and controls are superimposed. It is readily apparent that there are differences between cases and controls, and that the direction of the differences depends on which biomarker we consider. Both univariate summaries, $\kappa$ and AUC, signal a preference for the first biomarker as a screening mechanism. {The 95\% credible interval of $\kappa$ associated with Biomarker~1 is (0.71, 0.77) and for $\AUC$ is (0.80, 0.86), while for Biomarker~2 the interval for $\kappa$ is (0.83, 0.90)  and for $\AUC$  (0.70, 0.77) respectively. } 

On the one hand, total PSA (Biomarker~1) looks like a reasonable biomarker since $\AUC = 0.83$, but on the other hand $\kappa$ seems to be putting into question that evidence  ($\kappa = 0.74$). {The overlapping histograms in Figure~\ref{Figure:UnconditionalCancer} shed some light on the reasons underlying the lack of consensus between $\AUC$ and $\kappa$. Despite the fact that the AUC is moderately large in both cases, there is a considerable overlap between the distributions of the biomarkers for diseased and non-diseased subjects. And interestingly, the modes of both distributions are not that far apart. But perhaps this should not be regarded surprising since the discrimination power of PSA has been called into question and often regarded as controversial \citep{prensner2012, harvard2017}.}


\subsection{PSA-based analysis with age-adjustment}\label{condData}
It is well known that PSA levels may be age-dependent---for both diseased and non-diseased subjects---since both benign prostate conditions and prostate cancer become more common with age. With this in mind, we obtained conditional density estimates for each biomarker in each population to estimate $\kappa(\mbox{age})$ and $\AUC(\mbox{age})$ {by fitting the conditional model and collecting 1\,800 MCMC iterates after a burn in of 20\,000 and thinning of 100 and using the same specifications as before.} In model fitting, the patients' ages were first rescaled from the interval $[46.75, 80.83]$ to the interval $[-1,1]$, and, following numerical evidence from  \citet[][Section~3]{inaciodecarvalho2017}, we elected to not to include any additional knots in the cubic B-splines.

{Figure~\ref{condPSA} displays the posterior mean and pointwise 95\% credible intervals for $\kappa$ and AUC as a function of age. Notice first that for Biomarker~1 our estimated $\AUC(\mbox{age})$ is very similar to that found in Figure~4 of \cite{rodriguez2014}, with the largest discriminatory power occurring when an individual is in their late 50s.  Regarding comparisons with $\kappa$, generally speaking Biomarker~1  exhibits less affinity than Biomarker~2 between the distributions of those with and without a prostate cancer diagnosis. This suggests that a diagnostic test based on Biomarker~1 would be preferred to a test based on Biomarker~2. The first biomarker's affinity appears to be sensitive to the subject's age. In line with the findings in the previous section, AUC seems to indicate that PSA is a reasonably good diagnostic test, while $\kappa$ seems to be more pessimistic regarding the test's ability ($\kappa \approx 0.6$); a similar conclusion holds for Biomarker 2. In addition, $\kappa$ more clearly identifies the difference in screening ability of the two biomarkers for males aged 55 to 70. Furthermore, it is invariant to whether the diagnostic test is assumed to be lower- or upper-tailed.}
\begin{figure} \hspace{0.8cm}
  \begin{minipage}{0.275\linewidth} \hspace{-1.2cm}
    \includegraphics[scale = 0.4]{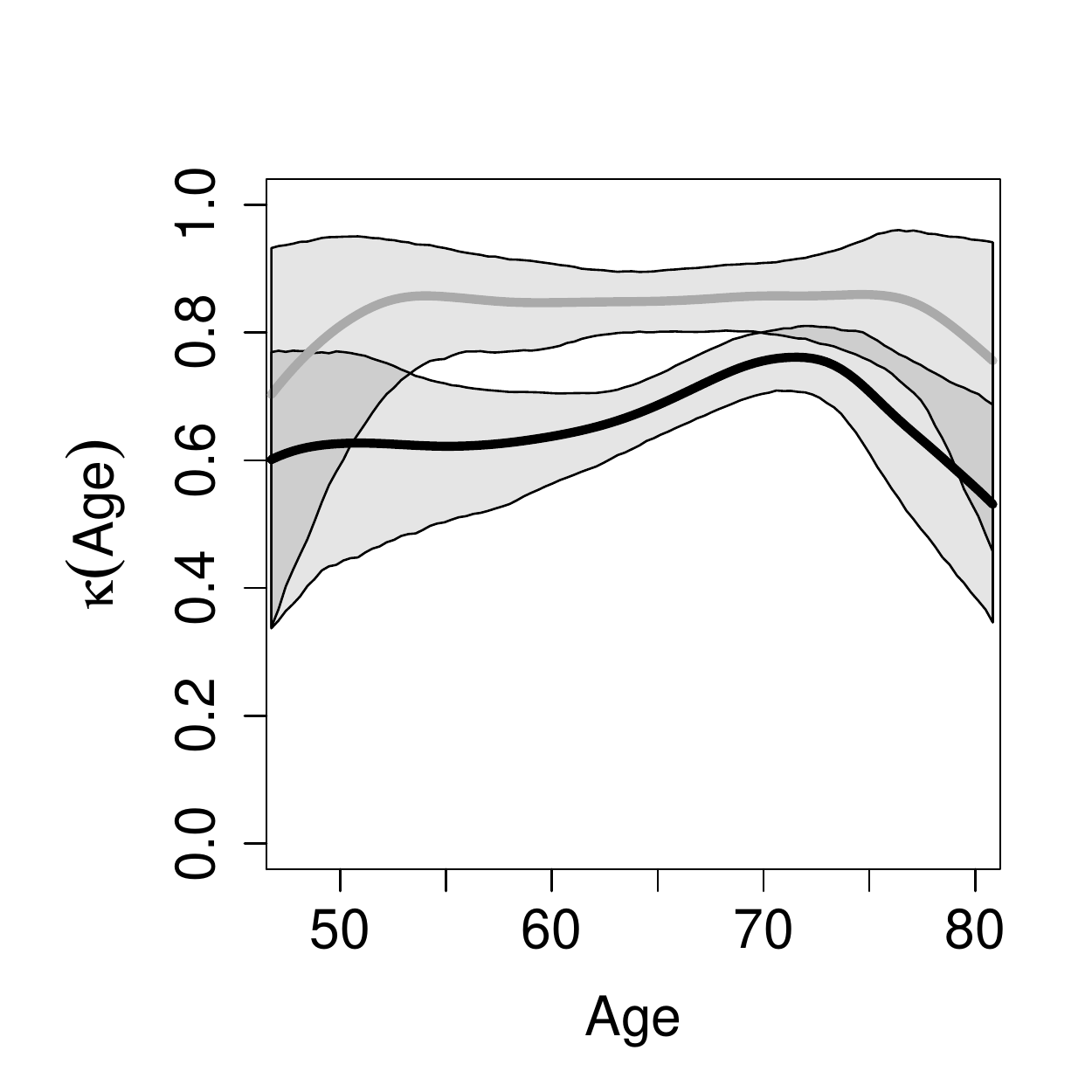}    
  \end{minipage} 
  \begin{minipage}{0.275\linewidth} \hspace{1cm}
    \includegraphics[scale = 0.4]{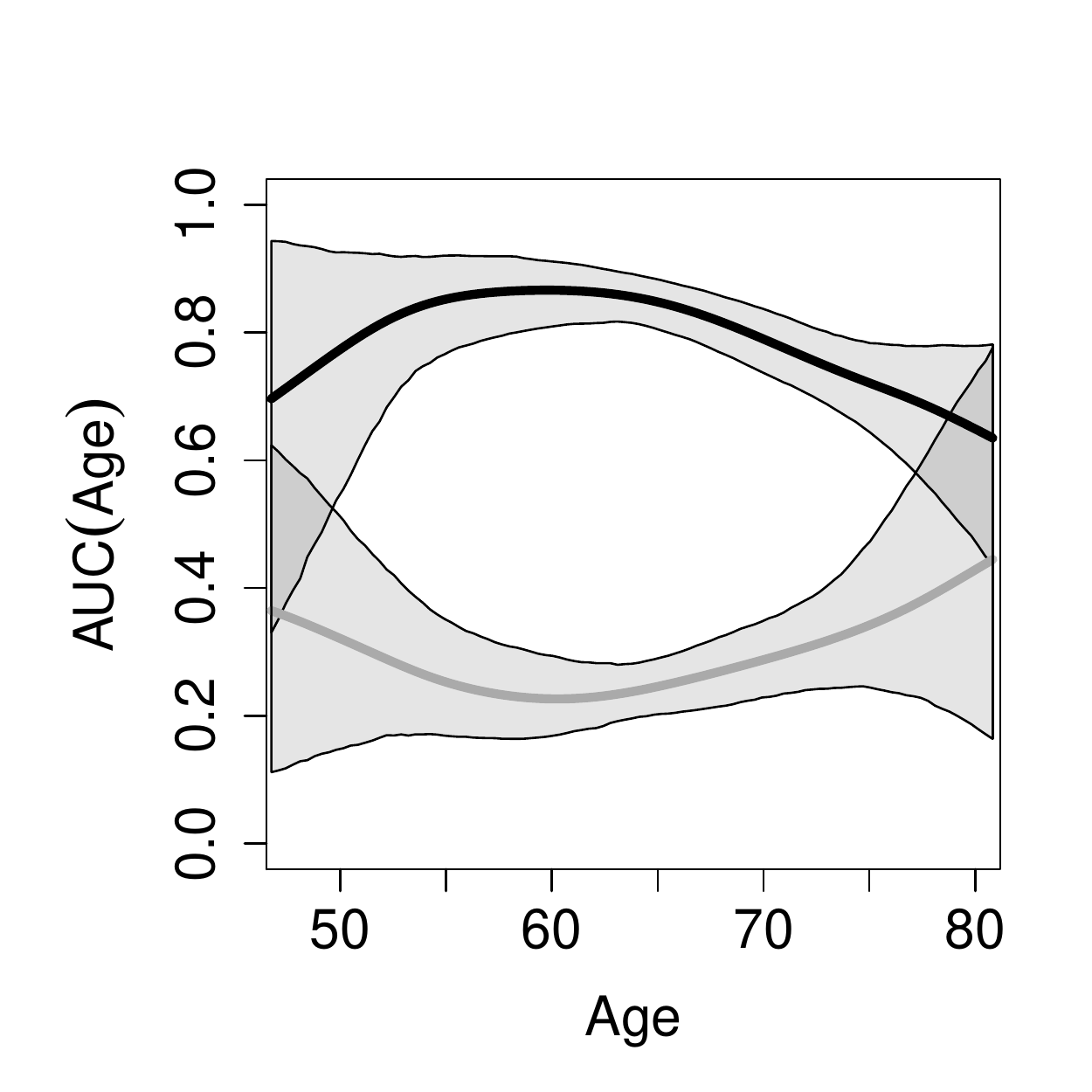}    
  \end{minipage} \hspace{1cm}
  \begin{minipage}{0.275\linewidth} \hspace{1cm}
    \includegraphics[scale = 0.4]{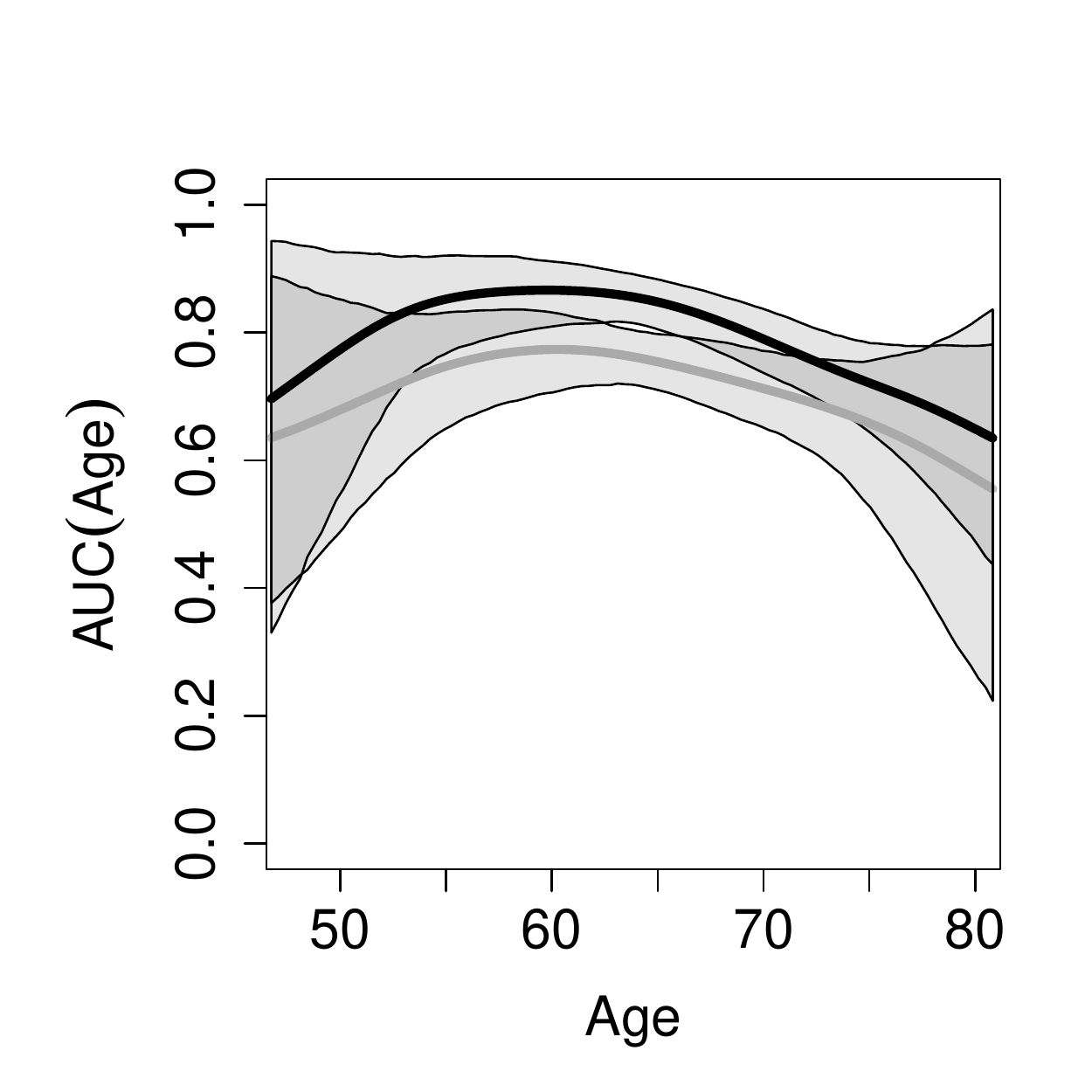}    
  \end{minipage} \\
  \begin{minipage}{0.275\linewidth} \centering \footnotesize
    ~~~~~~~(a)
  \end{minipage}\hspace{0.7cm}
  \begin{minipage}{0.275\linewidth} \centering \footnotesize
    ~~~~~~~~~~~~~~(b)
  \end{minipage}\hspace{0.7cm}
  \begin{minipage}{0.275\linewidth} \centering \footnotesize
    ~~~~~~~~~~~~~~~~~~~~~~(c)
  \end{minipage} 
    \caption{\small Means and 95\% pointwise credible intervals for the age-adjusted affinity and $\AUC$ of two biomarkers in cases and controls. a) is the age-adjusted affinity; b) is the age-adjusted $\AUC$ if both biomarkers have upper-tailed diagnostic tests; c) is the age-adjusted $\AUC$ if the second biomarker diagnostic test is lower-tailed. In each panel, the black and grey  lines respectively denote the first and second biomarkers.}
    \label{condPSA}
\end{figure}
Finally, both analyses suggest Biomarker~1 is a better alternative than Biomarker~2 in screening older males with lung cancer for prostate cancer. This latter conclusion is supported even more emphatically by $\kappa(\mbox{age})$ than by $\AUC(\mbox{age})$, as seen in Figure~\ref{condPSA}. 

\section{Discussion}
In this paper we show how Hellinger affinity can be used as a natural summary measure of medical diagnostic accuracy. The summary measure has several desirable properties that motivate its use as a supplement, if not competitor, to other existing summaries such as $\AUC$ and the Youden index. Affinity shares some of the properties of the AUC---such as invariance to monotone increasing transformations---, but it does not fall into the separation trap, {whereas both the AUC and the Youden index would}. Indeed, a principal advantage of $\kappa$ is that it is readily calculated and interpreted without assuming anything about the biomarker threshold(s) that demarcate positive and negative test diagnoses. This can be especially beneficial if, for instance, a biomarker's distribution when the disease is present favors both atypically low and atypically high values.   Affinity-based measures can be framed into the same geometrical principles as Pearson correlation, and they focus on the overlap between $f_D$ and $f_{\D}$ rather than on always presuming that larger values of {a biomarker} are more indicative of disease. Nonparametric Bayes estimators for affinity and covariate-specific affinity are discussed, and theoretical properties of the corresponding priors have been derived. While it could be natural to fit parametric models such as the ones in Table~\ref{parametric}, the added flexibility of the proposed inferences allows us to model diagnostic test accuracy in a way that offers flexibility and robustness against misspecification.

While not explored here, our summary measure has the potential to be applied to  the more general setting where $p > 1$ biomarkers per subject are available. Indeed, if $f_D(y)$ and $f_{\D}(y)$ denote the joint distributions of the $p$ biomarkers, for diseased and non-diseased subjects, similarly to \eqref{kappa} one can define  
\begin{equation*}
  \kappa = \langle \sqrt{f_D}, \sqrt{f_{\D}} \rangle  
  = \int_{\mathbb{R}^p} \sqrt{f_D(y)} \sqrt{f_{\D}(y)} \, \dif y.
\end{equation*}
{However, estimation of $\kappa$ would become more challenging in the multivariate case than in the univariate case presented in this article.} Future work could also entail nonparametric Bayesian inference for a covariate-specific version of the so-called overlap coefficient \citep{wang2017} which can be defined as 
\begin{equation*}
  \mbox{OVL}(\bmx) = \int_{-\infty}^{\infty} \min\{f_{D}(y \mid \bmx), f_{\D}(y \mid \bmx) \}\, \dif y.
  \label{OVL}
\end{equation*}
The index in \eqref{OVL} would quantify the proportion of overlap area between $f_{D}(y \mid \bmx)$ and $f_{\D}(y \mid \bmx)$, and thus it could be used as a companion to $\AUC(\bmx)$ and $\kappa(\bmx)$. Finally, another direction which we may revisit in future work rests on the study of $\kappa(\bmx)$ and $\mbox{OVL}(\bmx)$ on settings where a gold standard test is unavailable.  


\section*{Funding}
This work was partially supported by FCT (Funda\c c\~ao para a Ci\^encia e a Tecnologia, Portugal), through the project UID/MAT/00006/2013.
\bibliographystyle{biorefs}
\bibliography{bib.bib}

\section*{Supplementary materials}
\label{Supplementary Material}
Supplementary materials include proofs of theoretical results, derivations for entries in Table~\ref{parametric}, simulation setting figures, and a supplement to the data analysis. 

\section*{\S 1.~Properties of affinity}
\subsection*{\S 1.1.~Proof of Proposition~1}
  The proof is as follows:
  \begin{enumerate}
  \item Since $\sqrt{f_D(y)} \geqslant 0$ and $\sqrt{f_{\D}(y)} \geqslant 0$ it follows that $\kappa \geqslant 0$. Cauchy--Schwarz inequality further implies that $|\kappa| = \langle \sqrt{f_D},  \sqrt{f_{\D}} \rangle \leqslant \|\sqrt{f_D}\| \|\sqrt{f_{\D}}\| = 1$.
  \item The argument is tantamount to that of \cite{roos2011}, but it is included here for completeness. Let $g(y) = z$ be a monotone increasing function and let 
    \begin{equation*}
      f_D^g(z) = f_D(g^{-1}(z))\frac{\dif}{\dif z} g^{-1}(z), 
      \quad f_{\D}^g(z) = f_{\D}(g^{-1}(z))\frac{\dif}{\dif z} g^{-1}(z),      
    \end{equation*}
    be densities of the transformed data $g(Y_D)$ and $g(Y_{\D})$ \citep{knight2000}, and denote by $\kappa^g = \langle \sqrt{\mathstrut  f_D^g}, \sqrt{\mathstrut f_{\D}^g}\rangle$ the affinity of the transformed data. It thus follows that  
    \begin{equation*}
      \begin{split}
        \kappa^g &= \int_{-\infty}^{\infty} \sqrt{f_D^g(z)}\sqrt{f_{\D}^g(z)} \,\dif z \\
        &= \int_{-\infty}^{\infty} \left\{f_D(g^{-1}(z))\frac{\dif}{\dif z} g^{-1}(z)\right\}^{1/2} \left\{f_{\D}(g^{-1}(z))\frac{\dif}{\dif z} g^{-1}(z)\right\}^{1/2} \, \dif z \\
& = \int_{-\infty}^{\infty} \sqrt{f_D(g^{-1}(z))}\sqrt{f_{\D}(g^{-1}(z))} \frac{\dif}{\dif z} g^{-1}(z) \, \dif z \qquad (\text{set } g^{-1}(z) = y) \\
& = \int_{-\infty}^{\infty} \sqrt{f_D(y)}\sqrt{f_{\D}(y)}\, \dif y = \kappa.
      \end{split}
    \end{equation*}
  \end{enumerate}

\section*{\S 2.~Properties of induced priors}

\subsection*{\S 2.1.~Proof of Theorem~1}
The proof involves some manipulations similar to those often used for showing continuity of the inner product \citep{hunter2001}, along with a result from  \cite{lijoi2004}. Just note that
  \begin{equation}
    \begin{split}
      |\kappa^{\omega} - \kappa| &= 
      |\langle \sqrtfdw, \sqrtfndw \rangle - 
      \langle \sqrtfd, \sqrtfnd \rangle|\\ 
      &= |\langle \sqrtfdw, \sqrtfndw \rangle - 
      \langle \sqrtfdw, \sqrtfnd \rangle + 
      \langle \sqrtfdw, \sqrtfnd \rangle 
      - \langle \sqrtfd, \sqrtfnd \rangle| \\
      &\leqslant |\langle \sqrtfdw, \sqrtfndw \rangle - 
      \langle \sqrtfdw, \sqrtfnd \rangle| + 
      |\langle \sqrtfdw, \sqrtfnd \rangle 
      - \langle \sqrtfd, \sqrtfnd \rangle| \\
      &\leqslant |\langle \sqrtfdw, \sqrtfndw - \sqrtfnd \rangle| +         
      |\langle \sqrtfdw - \sqrtfd, \sqrtfnd \rangle|\\
      &\leqslant \underbrace{\|\sqrt{f_{D}^\omega}\|}_1 \|\sqrt{f_{\D}^\omega} - \sqrt{f_{\D}}\| + 
      \|\sqrt{f_{D}^\omega} - \sqrt{f_{D}}\| \underbrace{\|\sqrt{f_{\D}}\|}_1\\
      &= \sqrt{d_{H}(f_{\D}^\omega, f_{\D})} + \sqrt{d_{H}(f_{D}^\omega, f_{D})}, 
      \label{manip}
    \end{split}
  \end{equation}
  where $d_{H}(f, g) = \int \{\sqrt{f(y)} - \sqrt{g(y)}\}^2 \, \dif y$ is the Hellinger distance. So, as it can be seen from \eqref{manip}, to have $|\kappa^{\omega} - \kappa| < \varepsilon$, with $\varepsilon > 0$, it would suffice having $d_{H}(f_{\D}^\omega, f_{\D}) < \varepsilon^2 / 4$ and $d_{H}(f_{D}^\omega, f_{D}) < \varepsilon^2 / 4$. Thus, 
  \begin{equation*}
    \{\omega \in \Omega: |\kappa^{\omega} - \kappa| < \varepsilon\} 
    \supseteq
    \{\omega \in \Omega: d_{H}(f_{\D}^\omega, f_{\D}) < \varepsilon^2 / 4, 
    d_{H}(f_{D}^\omega, f_{D}) < \varepsilon^2 / 4\},
  \end{equation*}
from where it finally follows that 
\begin{equation*}
  \begin{split}
    P\{\omega \in \Omega: |\kappa^{\omega} - \kappa| < \varepsilon\}
    \geqslant \underbrace{P\{\omega \in \Omega: d_{H}(f_{\D}^\omega,
      f_{\D}) < \varepsilon^2 / 4\}}_{\pi_{\D}} \underbrace{P\{\omega
      \in \Omega: d_{H}(f_{D}^\omega, f_{D}) < \varepsilon^2 /
      4\}}_{\pi_{D}} > 0,
  \end{split}
\end{equation*}
given that from the equivalence between Hellinger and $L_1$ support, and from Section~3 in \cite{lijoi2004} it follows that $\pi_{\D} > 0$ and $\pi_{D} > 0$.

\subsection*{\S 2.2.~Proof of Theorem~2}
The proof is along the same lines as that of Theorem~1 but it requires Theorem~4 of \cite{barrientos2012}, which is essentially a covariate-specific version of the result in Section~3 of \cite{lijoi2004}. Similar derivations as those in \eqref{manip} yield 
\begin{equation*}
  |\kappa^{\omega}(\bmx_i) - \kappa(\bmx_i)| \leqslant 
  \sqrt{d_H(f_{\D|\bmx_i}^\omega, f_{\D|\bmx_i})} + 
  \sqrt{d_H(f_{D|\bmx_i}^\omega, f_{D|\bmx_i})}.
\end{equation*}
Hence, to have $|\kappa^{\omega}(\bmx_i) - \kappa(\bmx_i)| < \varepsilon$, for $\varepsilon > 0$, it would suffice having 
\begin{equation*}
d_{H}(f_{\D|\bmx_i}^\omega, f_{\D|\bmx_i}) < \varepsilon^2 / 4, \quad 
d_{H}(f_{D|\bmx_i}^\omega, f_{D|\bmx_i}) < \varepsilon^2 / 4, \quad 
i = 1, \ldots, n,
\end{equation*}
and thus using similar arguments to the ones in proof of Theorem~1 it follows that 
\begin{gather*}
  \begin{split}
    P\{\omega \in \Omega: |\kappa^{\omega}(\bmx_i) - \kappa(\bmx_i)| <
    \varepsilon\} &\geqslant \underbrace{P\{\omega \in \Omega:
      d_{H}(f_{\D|\bmx_i}^\omega, f_{\D|\bmx_i}) < \varepsilon^2 / 4, i =
      1, \ldots, n\}}_{\pi_{\D}} \\ 
&\hspace{1cm}\times \underbrace{P\{\omega \in \Omega:
      d_{H}(f_{D|\bmx_i}^\omega, f_{D|\bmx_i}) < \varepsilon^2 / 4, i = 1,
      \ldots, n\}}_{\pi_D} > 0,
  \end{split}
\end{gather*}
given that Theorem~4 of \cite{barrientos2012}, on the Hellinger support of the DDP, implies that $\pi_{\D} > 0$ and $\pi_D > 0$.

\section*{\S 3.~Simulation setting figures}
Because of limited space in the article, only the second unconditional setting (70/30 mixtures of normals) from the simulation study was illustrated with a figure. We therefore include Figure~\ref{simsetu1} to display the densities, $\kappa$s, and AUCs from the first unconditional setting of Table~2. We also had included three covariate-dependent settings. Due to the challenges inherent in overlaying bivariate densities, we do not display the densities for the three conditional settings of Table~2, instead displaying only $\kappa(x)$ and AUC$(x)$. These covariate-dependent quantities are plotted in Figure~\ref{CondSettings}. 

\begin{figure}\centering
    \includegraphics[scale = 0.72]{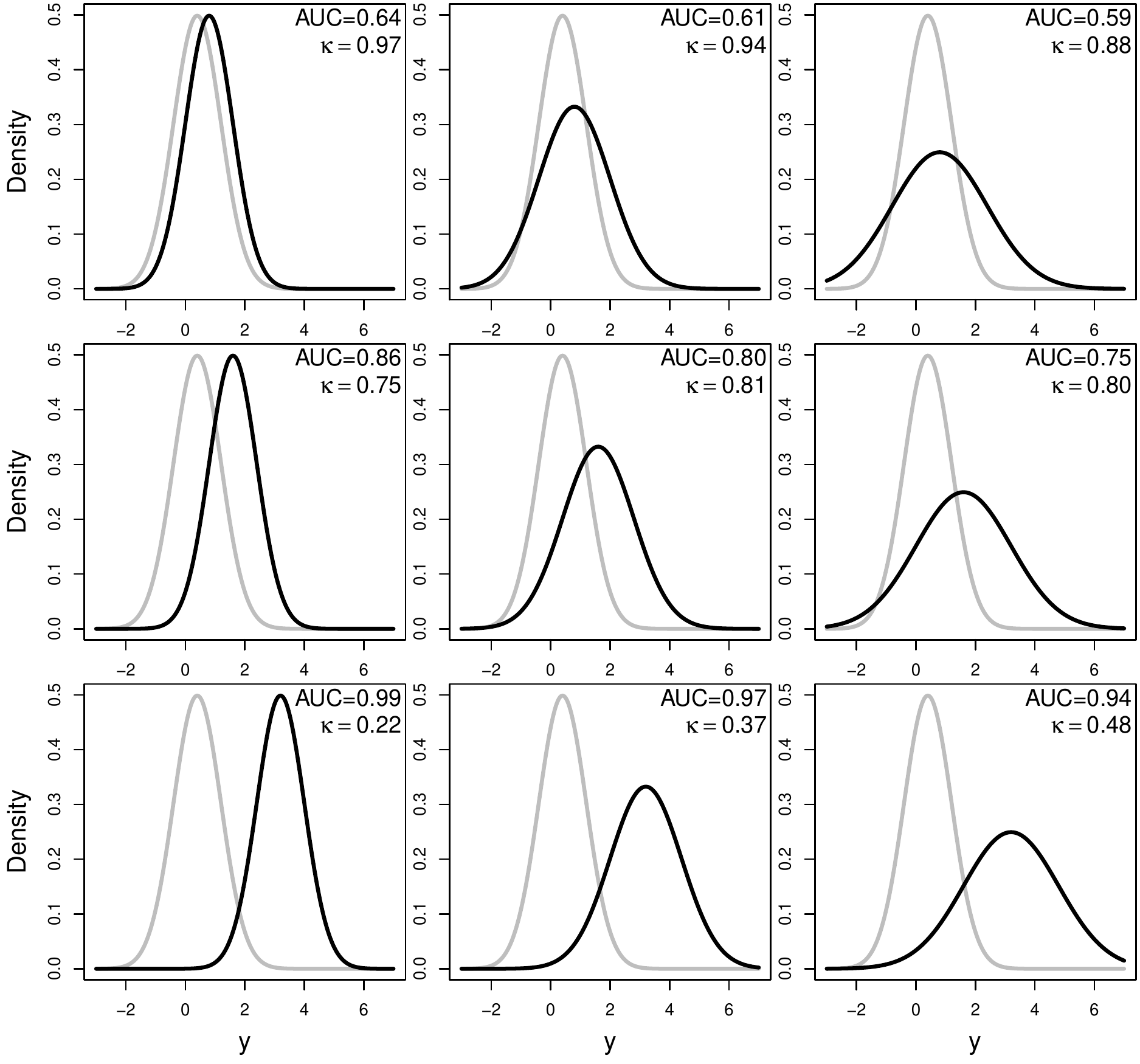}
   \caption[Distributions from Unconditional Setting 1]{
     Densities for the first unconditional simulation study 
     setting in Table~2 of the article; the black and grey lines respectively denote the densities of the biomarkers of the diseased and non-diseased subjects.}
    \label{simsetu1}
\end{figure}

\begin{figure}\centering
    \includegraphics[scale = 0.48]{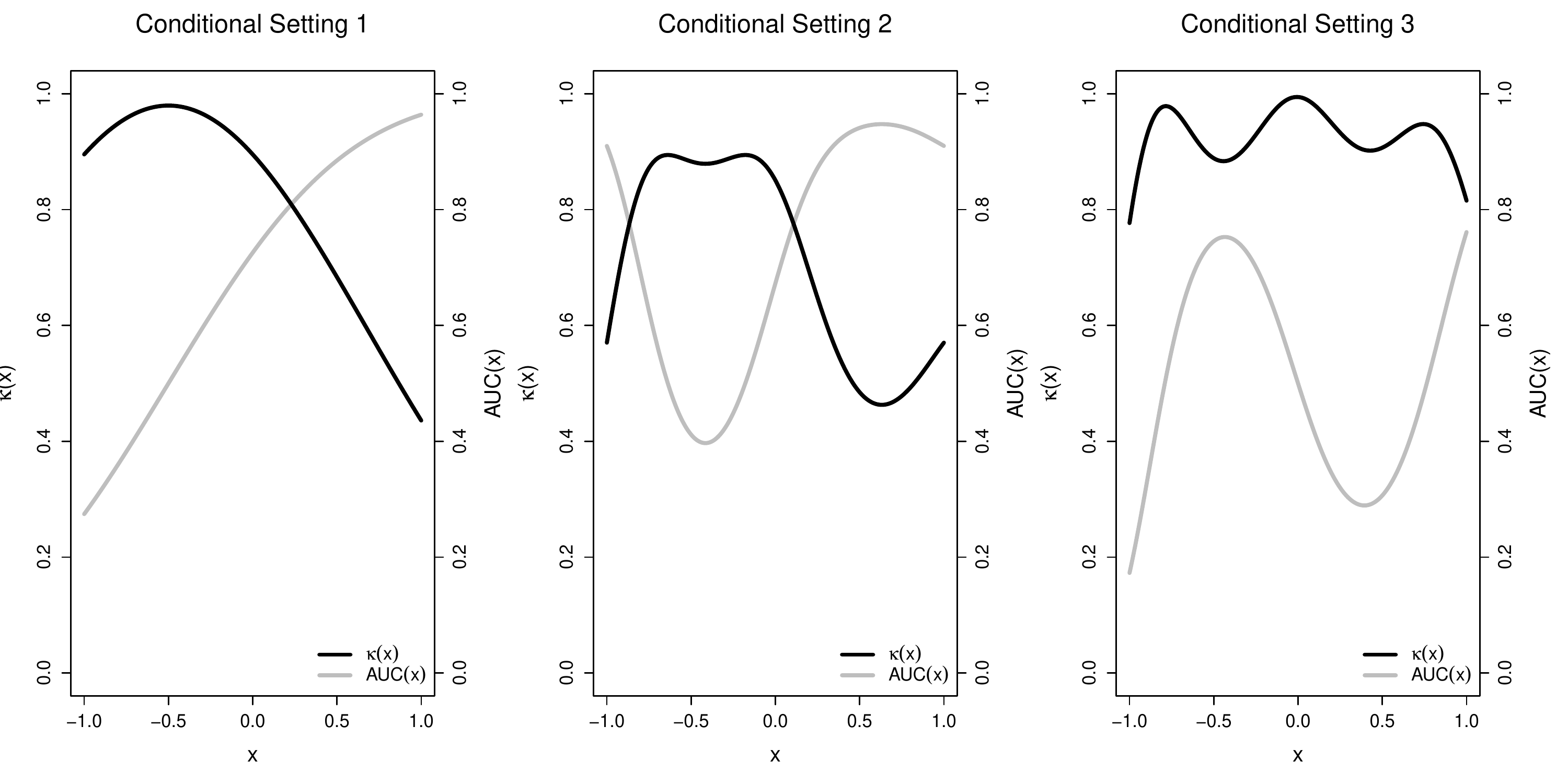}
   \caption[Diagnostic Accuracy Measures from Conditional Settings]{
     Covariate-dependent affinity and AUC for the three conditional settings 
     in Table~2 of the article; the black and grey lines respectively denote $\kappa(x)$ and AUC$(x)$.}
    \label{CondSettings}
\end{figure}

\section*{\S 4.~Additional empirical reports}
In the PSA data application (as reported in the article), the analysis proceeded as though 683 independent measures were obtained. This was done to enable comparison with an existing analysis of this data set by \cite{rodriguez2014}. One natural alternative is to use only the last available observation from each of the $n=141$ subjects. We made this restriction and repeated the analyses. In doing so, the range of observed ages was slightly reduced from [46.75, 80.83]---the range with the 683 observations---to [51.94, 80.83]. 
To apply the model in light of this smaller age range, the ages were rescaled so that 51.94 became $x=-1$ and 80.83 became $x=1$.  

The results from the unconditional and age-dependent analyses are contained in Figure~\ref{Figure:UncondPSA_Last} and Figure~\ref{Figure:CondPSA_Last}, respectively. The results when restricting the data to the last available observations are rather similar to the results when ignoring the dependence among all 683 observations. However, not only is there more uncertainty in the affinity and AUC estimates, but the estimates of both $\kappa$ and $\AUC$ are slightly more favorable with the 141 observations than they were with the 683 observations. This might be due to the fact that the last available observations on the PSA biomarkers were the closest observations to the time of diagnosis.

\begin{figure}[H]\centering
\vspace{-1.5cm}
\begin{center}
\includegraphics[scale=0.5]{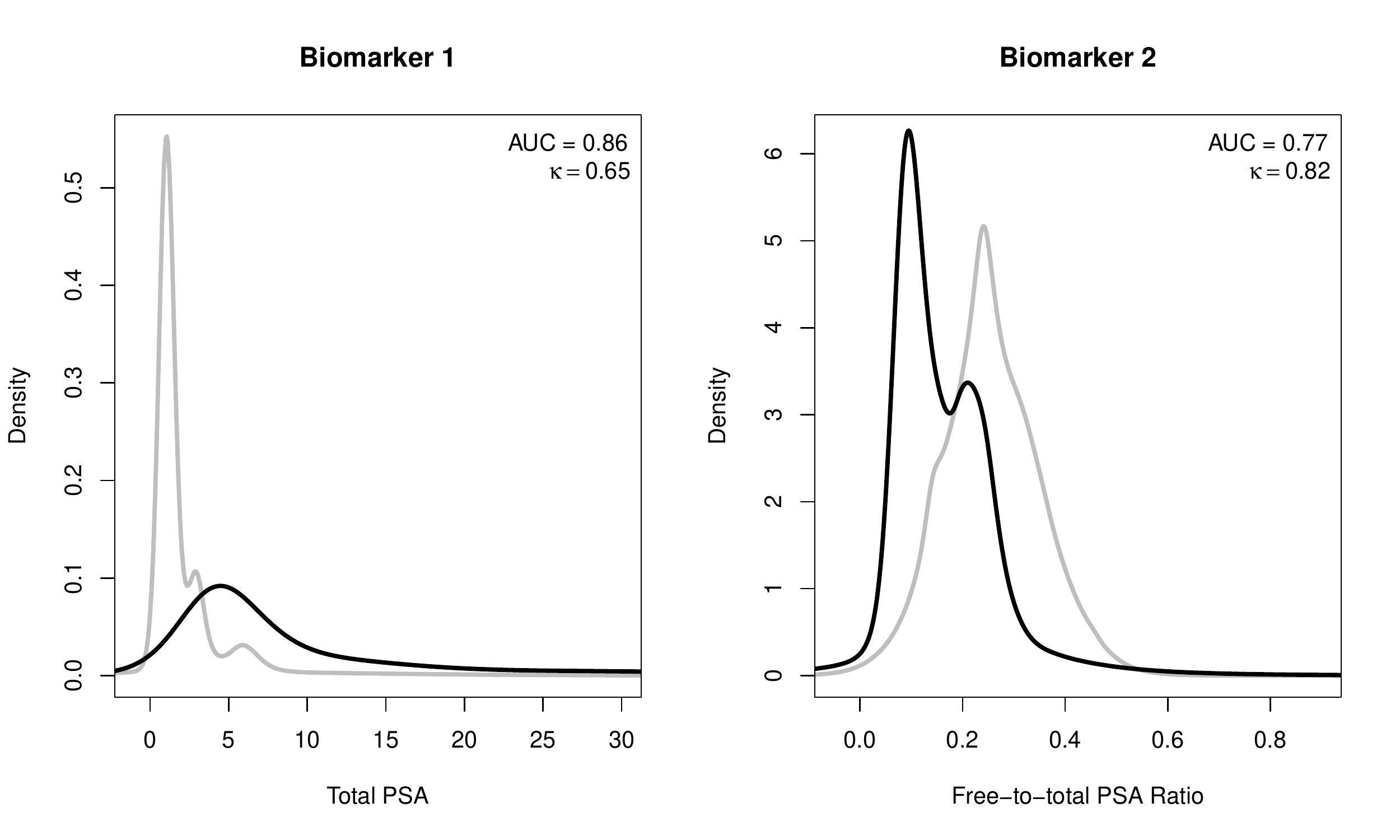}\\ \vspace{-4.2cm}
\includegraphics[scale=0.7]{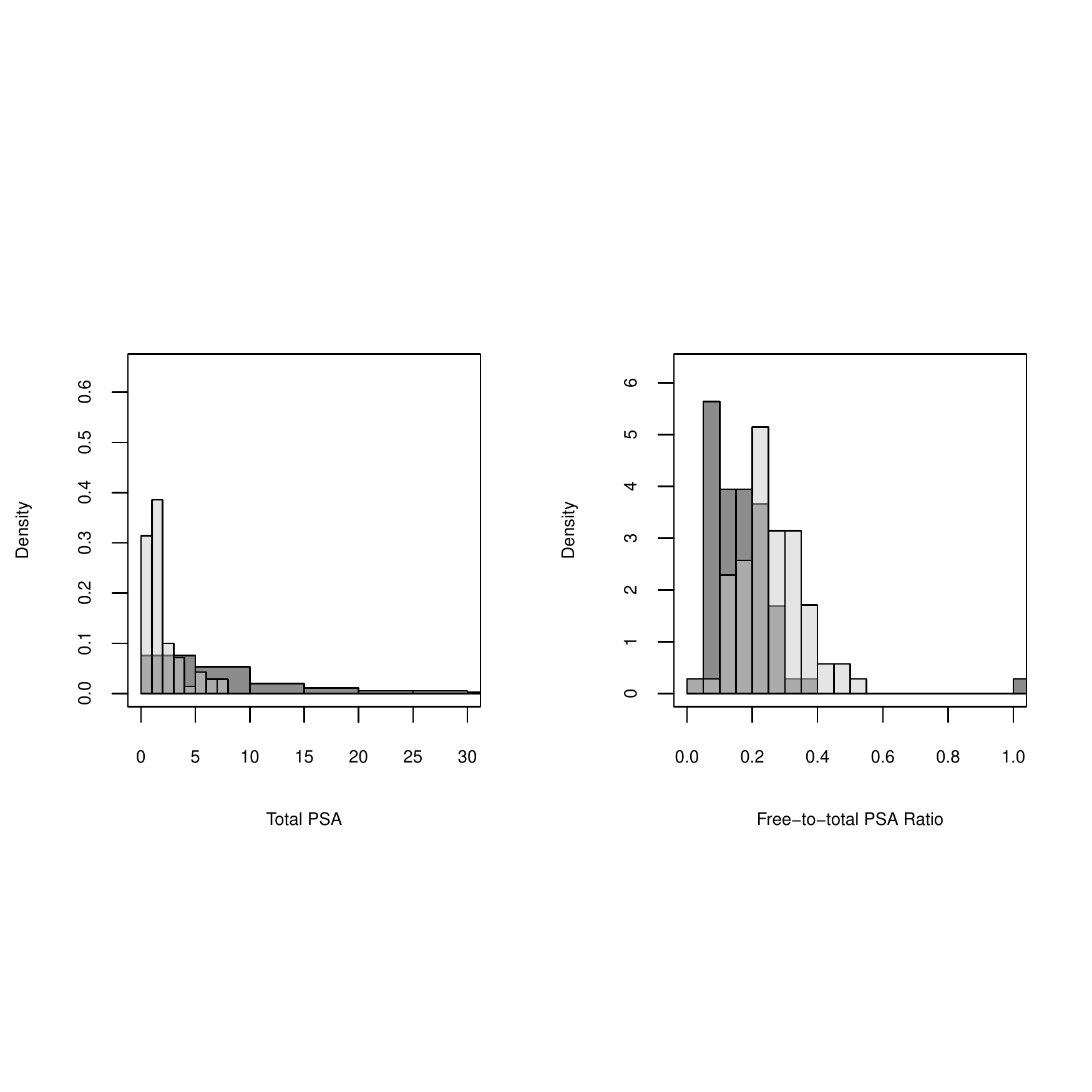} \\  \vspace{-3.3cm}
\caption{Last available observation analysis~I; this figures compares with Fig.~7 on the manuscript. Top: DPM-based estimated densities along with AUC and $\kappa$ values when age is not considered.  The black and grey lines respectively denote the densities of the biomarkers of the diseased and non-diseased subjects. Bottom: Overlapping histograms.}
\label{Figure:UncondPSA_Last}
\end{center}
\end{figure}
\begin{figure}[H]\hspace{0.8cm}
  \begin{minipage}{0.275\linewidth} \hspace{-1.2cm}
    \includegraphics[scale = 0.4]{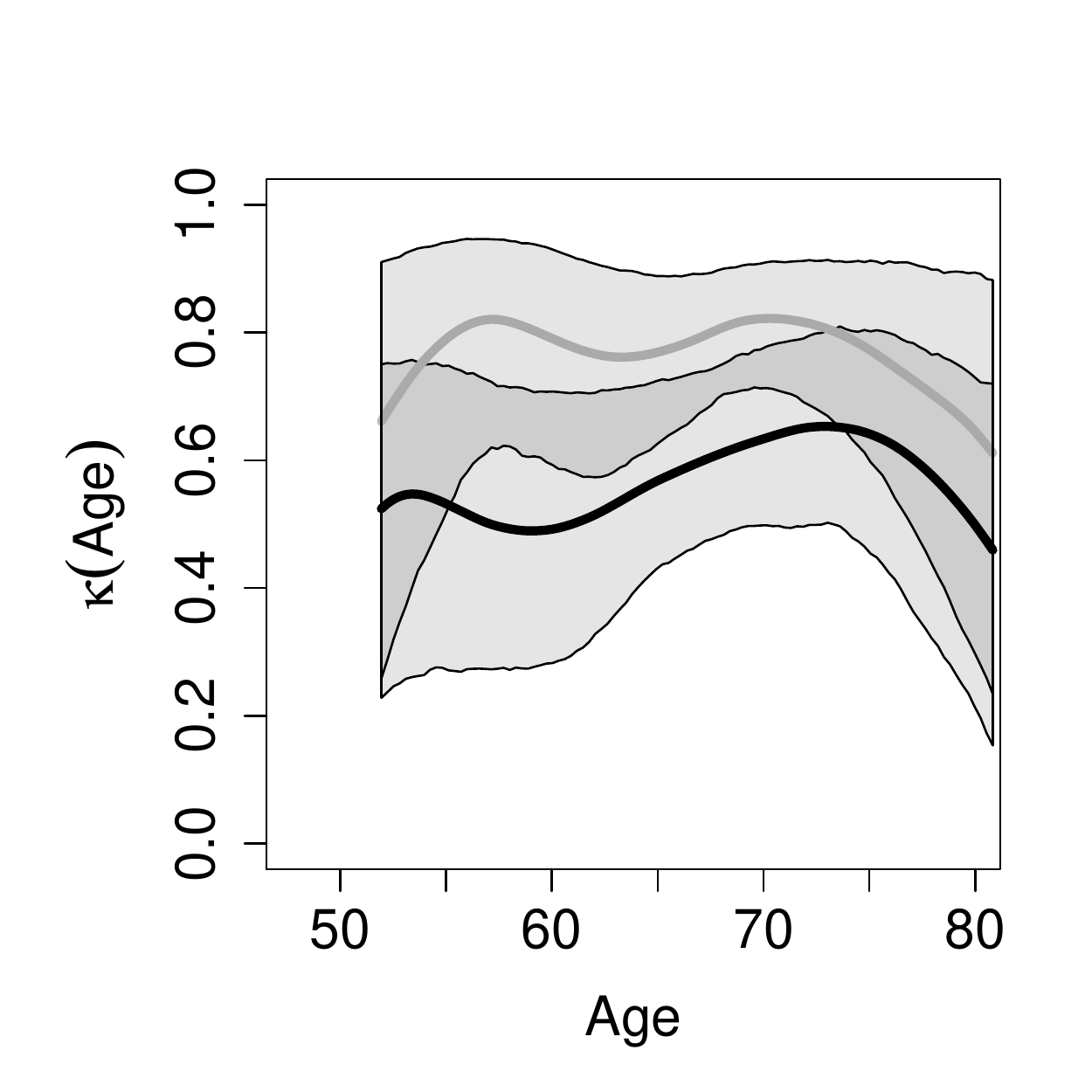}    
  \end{minipage}
  \begin{minipage}{0.275\linewidth} \hspace{1cm}
    \includegraphics[scale = 0.4]{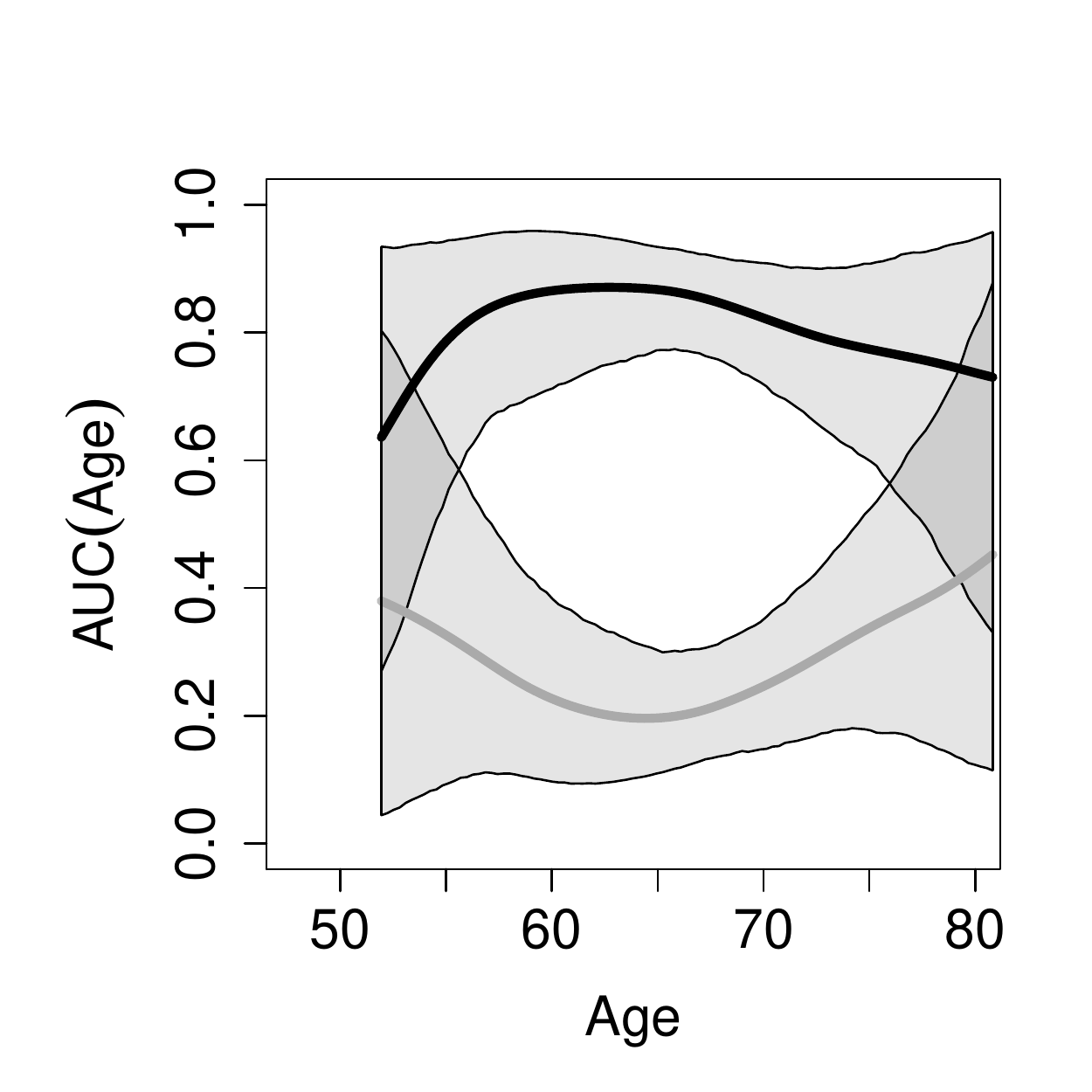}    
  \end{minipage}\hspace{1cm}
  \begin{minipage}{0.275\linewidth} \hspace{1cm}
    \includegraphics[scale = 0.4]{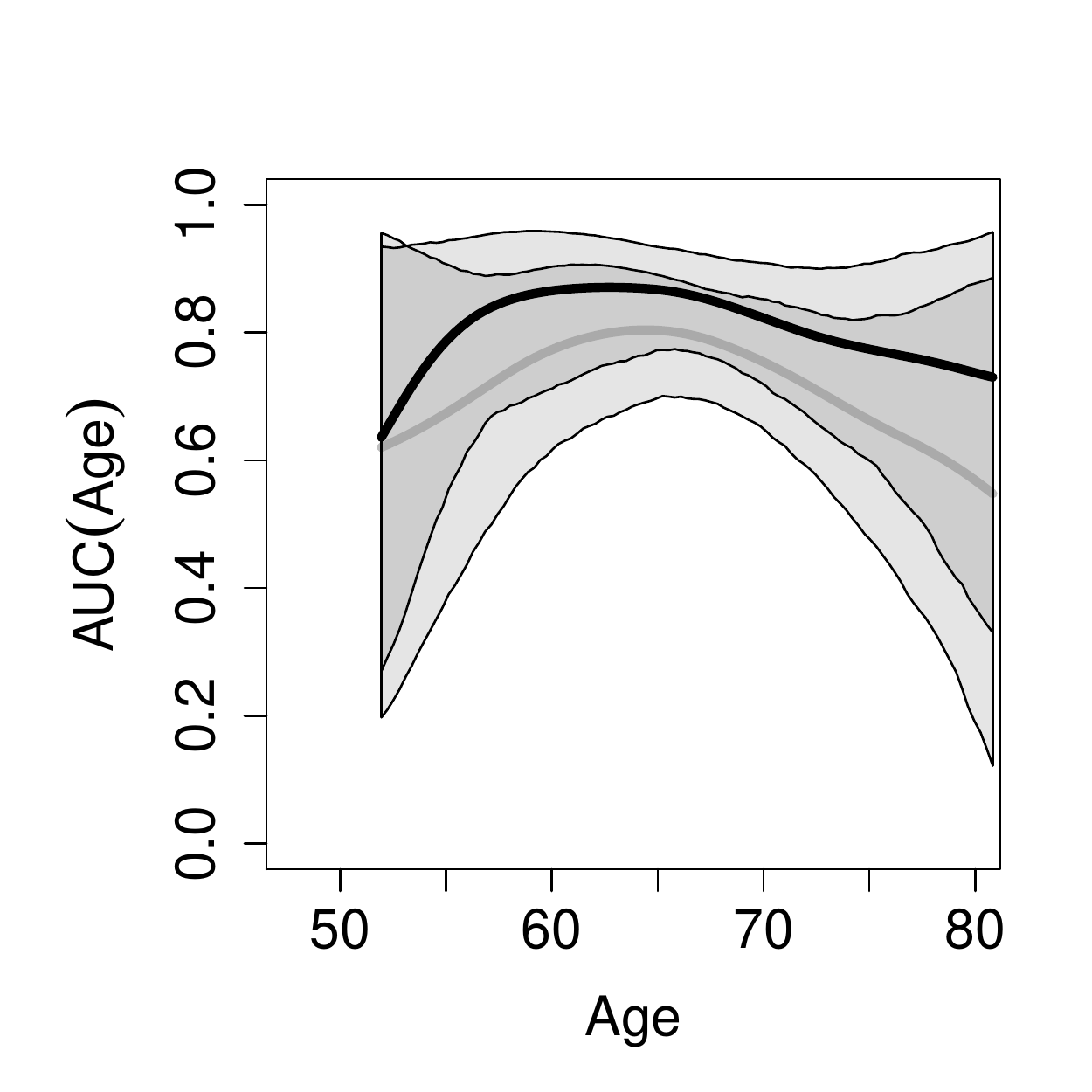}    
  \end{minipage} \\
  \begin{minipage}{0.275\linewidth} \centering \footnotesize
    ~~~~~~~(a)
  \end{minipage}\hspace{0.7cm}
  \begin{minipage}{0.275\linewidth} \centering \footnotesize
    ~~~~~~~~~~~~~~(b)
  \end{minipage}\hspace{0.7cm}
  \begin{minipage}{0.275\linewidth} \centering \footnotesize
    ~~~~~~~~~~~~~~~~~~~~(c)
  \end{minipage} 
    \caption{Last available observation analysis~II; this figures compares with Fig.~8. Means and 95\% pointwise credible intervals for the age-adjusted affinity and $\AUC$ of two biomarkers in cases and controls. Only the last available observation per subject was considered. a) is the age-adjusted affinity; b) is the age-adjusted $\AUC$ if both biomarkers have upper-tailed diagnostic tests; c) is the age-adjusted $\AUC$ if the second biomarker diagnostic test is lower-tailed. In each panel, the black and grey lines respectively denote the first and second biomarkers.}
    \label{Figure:CondPSA_Last}
\end{figure}

\section*{\S 5.~Analytical derivations of entries in Table~1}
\subsection*{Bibeta}
Let $Y_D \sim \mbox{Beta}(a_D, b_D)$ and $Y_{\D} \sim \mbox{Beta}(a_{\D}, b_{\D})$. Then,
\begin{equation*}
\begin{split}
\kappa &= \int_{-\infty}^{+\infty}{\sqrt{f_D(y)} \sqrt{f_{\D}(y)} \, \dif y}\\
& = \int_{0}^{1}{ \bigg\{\frac{y^{a_{D} - 1}(1-y)^{b_{D}-1}}{B(a_D, b_D)}\bigg\}^{1/2} \bigg\{\frac{y^{a_{\D} - 1}(1-y)^{b_{\D}-1}}{B(a_{\D}, b_{\D})}\bigg\}^{1/2} \, \dif y} \\
& =\frac{\int_0^1{ y^{(a_D + a_{\D})/2 -1} (1-y)^{(b_D + b_{\D})/2 -1}\, \dif y}}{\{B(a_D, b_D)B(a_{\D}, b_{\D})\}^{1/2}} 
 = \frac{B((a_D +a_{\D}) / 2, (b_D + b_{\D}) / 2)}{\{B(a_D, b_D)B(a_{\D}, b_{\D})\}^{1/2}},
\end{split}
\end{equation*}
where $B(a, b) = \int_0^1 u^{a - 1} (1 - u)^{b - 1} \, \dif u$.

\subsection*{Biexponential}
Let $Y_D \sim \Exp(\lambda_D)$ and $Y_{\D} \sim \Exp(\lambda_{\D})$. Then,
\begin{equation*}
\begin{split}
\kappa &= \int_{-\infty}^{+\infty}{\sqrt{f_D(y)} \sqrt{f_{\D}(y)} \, \dif y}
 = \int_{0}^{+\infty}{ \sqrt{\lambda_D \exp\{ -\lambda_D y \} } \sqrt{\lambda_{\D} \exp\{-\lambda_{\D} y \} )} \, \dif y} \\
& = (\lambda_D\lambda_{\D})^{1/2} \int_0^{+\infty}{\exp\bigg\{ - \frac{\lambda_D + \lambda_{\D}}{2} \, y \bigg\} \, \dif y} = \frac{2(\lambda_D \lambda_{\D})^{1/2}}{\lambda_D + \lambda_{\D}}. 
\end{split}
\end{equation*}

\subsection*{Binormal}
Let $Y_D \sim \mbox{N}(\mu_D$, $\sigma_D)$ and 
$Y_{\D} \sim \mbox{N}(\mu_{\D}$, $\sigma_{\D})$. Then, 
\begin{gather*}
\begin{split}
\kappa &= \int_{-\infty}^{+\infty}{\sqrt{f_D(y)} \sqrt{f_{\D}(y)}\, \dif y}\\
& = \int_{-\infty}^{+\infty}{ \bigg[ (2\pi\sigma_{D}^2)^{-1/2} \exp\bigg\{-\frac{1}{2}\frac{(y-\mu_D)^2}{\sigma_D^2} \bigg\} \bigg]^{1/2}  \bigg[(2\pi\sigma_{\D}^2)^{-1/2} \exp\bigg\{-\frac{1}{2}\frac{(y-\mu_{\D})^2}{\sigma_{\D}^2} \bigg\} \bigg]^{1/2}\, \dif y} \\
& = \frac{1}{\sqrt{2\pi\sigma_{D}\sigma_{\D}}} \int_{-\infty}^{+\infty}{\exp \bigg\{  -\frac{1}{4}\bigg(  \frac{(y-\mu_D)^2}{\sigma_D^2} + \frac{(y-\mu_{\D})^2}{\sigma_{\D}^2} \bigg)  \bigg\}\, \dif y} \\
& = \frac{1}{\sqrt{2\pi\sigma_{D}\sigma_{\D}}} \int_{-\infty}^{+\infty}{\exp \bigg\{  -\frac{1}{4\sigma_D^2 \sigma_{\D}^2} \bigg(  (\sigma_D^2 + \sigma_{\D}^2)y^2 - 2(\sigma_{\D}^2\mu_D + \sigma_D^2\mu_{\D})y + \sigma_{\D}^2\mu_D^2 + \sigma_D^2\mu_{\D}^2 \bigg)  \bigg\}\, \dif y} \\
& = \frac{1}{\sqrt{2\pi\sigma_{D}\sigma_{\D}}}  \exp \bigg\{  -\frac{1}{4 \sigma_D^2\sigma_{\D}^2 } \bigg(
\sigma_{\D}^2 \mu_D^2 + \sigma_D^2 \mu_{\D}^2   -   \frac{(\sigma_{\D}^2\mu_D + \sigma_D^2 \mu_{\D})^2}{\sigma_{\D}^2 + \sigma_D^2}  \bigg) 
    \bigg\} \\ & \hspace{2cm} \times\int_{-\infty}^{+\infty}{  \exp \bigg\{  -\frac{1}{2 (2\sigma_D^2\sigma_{\D}^2/(\sigma_D^2 + \sigma_{\D}^2)) } \bigg(
y- \bigg( \frac{\sigma_{\D}^2\mu_D + \sigma_D^2 \mu_{\D}}{\sigma_{\D}^2 + \sigma_D^2} \bigg) \bigg)^2 
    \bigg\}\, \dif y} \\
& =  \frac{1}{\sqrt{2\pi\sigma_{D}\sigma_{\D}}}  \sqrt{2 \pi \frac{2\sigma_D^2\sigma_{\D}^2}{\sigma_D^2 + \sigma_{\D}^2}}  \exp \bigg\{  -\frac{1}{4 \sigma_D^2\sigma_{\D}^2 } \bigg(
\sigma_{\D}^2 \mu_D^2 + \sigma_D^2 \mu_{\D}^2 - \frac{(\sigma_{\D}^2\mu_D + \sigma_D^2 \mu_{\D})^2}{\sigma_{\D}^2 + \sigma_D^2}  \bigg) 
    \bigg\}  \\
& = \sqrt{\frac{2 \sigma_D \sigma_{\D}}{\sigma_D^2 + \sigma_{\D}^2}} \exp\bigg\{-\frac{1}{4}
    \frac{(\mu_D - \mu_{\D})^2}{\sigma_D^2 + \sigma_{\D}^2} \bigg\}.\\
\end{split}
\end{gather*}

\end{document}